\pdfoutput=1

\documentclass[iop]{emulateapj}

\usepackage{amsmath}
\usepackage{natbib}
\usepackage{mhchem}
\usepackage{chngcntr}
\usepackage{threeparttable}
\newcommand{\htoj}{$\mathrm{H}_2\mathrm{O} - \mathrm{J}$}
\newcommand{\teq}{$T_\mathrm{eq}$}
\newcommand{\tirr}{$T_\mathrm{irr}$}
\newcommand{\myemail}{gbruno@stsci.edu}
\def\gtsima{$\; \buildrel > \over \sim \;$}
\def\ltsima{$\; \buildrel < \over \sim \;$}
\def\gtrsim{\lower.5ex\hbox{\gtsima}}
\def\lesssim{\lower.5ex\hbox{\ltsima}}

\slugcomment{}

\shorttitle{WASP-67\lowercase{b} and HAT-P-38\lowercase{b}}
\shortauthors{Bruno et al.}

\begin{document}


\title{A comparative study of WASP-67\lowercase{b} and HAT-P-38\lowercase{b} from WFC3 data}


\author{Giovanni Bruno\altaffilmark{1}, Nikole K. Lewis, Kevin B. Stevenson, Joseph Filippazzo, Matthew Hill, Jonathan D. Fraine, Hannah R. Wakeford}
\affil{Space Telescope Science Institute, 3700 San Martin Drive, Baltimore, MD 21218, USA}

\author{Drake Deming}
\affil{Department of Astronomy, University of Maryland at College Park, College Park, MD 20742, USA}

\author{Brian Kilpatrick}
\affil{Department of Physics, Box 1843, Brown University, Providence, RI 02904, USA}

\author{Michael R. Line}
\affil{School of Earth \& Space Exploration, Arizona State University, Phoenix, AZ 85282, USA}

\author{Caroline V. Morley}
\affil{Department of Astronomy, Harvard University, Cambridge, MA 02138, USA}

\author{Karen A. Collins}
\affil{Department of Physics and Astronomy, Vanderbilt University, Nashville, TN 37235, USA}

\author{Dennis M. Conti}
\affil{American Association of Variable Star Observers, 49 Bay State Rd., Cambridge, MA 02138 USA}

\author{Joseph Garlitz}
\affil{1155 Hartford St.Elgin, Oregon 97827 USA}

\author{Joseph E. Rodriguez}
\affil{Harvard-Smithsonian Center for Astrophysics, 60 Garden Street, Cambridge, MA 02138, USA}


\altaffiltext{1}{\myemail}


\begin{abstract}
Atmospheric temperature and planetary gravity are thought to be the main parameters affecting cloud formation in giant exoplanet atmospheres. Recent attempts to understand cloud formation have explored wide regions of the equilibrium temperature-gravity parameter space. In this study, we instead compare the case of two giant planets with nearly identical equilibrium temperature (\teq\ $\sim 1050 \, \mathrm{K}$) and gravity ($g \sim 10 \, \mathrm{m \, s}^{-1})$. During {\em HST} Cycle 23, we collected WFC3/G141 observations of the two planets, WASP-67 b and HAT-P-38 b. HAT-P-38 b, with mass 0.42 M$_\mathrm{J}$ and radius 1.4 $R_\mathrm{J}$, exhibits a relatively clear atmosphere with a clear detection of water. We refine the orbital period of this planet with new observations, obtaining $P = 4.6403294 \pm 0.0000055 \, \mathrm{d}$. WASP-67 b, with mass 0.27 M$_\mathrm{J}$ and radius 0.83 $R_\mathrm{J}$, shows a more muted water absorption feature than that of HAT-P-38 b, indicating either a higher cloud deck in the atmosphere or a more metal-rich composition. The difference in the spectra supports the hypothesis that giant exoplanet atmospheres carry traces of their formation history. Future observations in the visible and mid-infrared are needed to probe the aerosol properties and constrain the evolutionary scenario of these planets.
\end{abstract}





\section{Introduction}
Exoplanet atmospheres are a unique window into the composition of exoplanets, their physical-chemical characteristics, and the interaction with their host stars. The degeneracy between aerosols and metallicity prevents from precisely constraining atmospheric composition \citep[e.g.][]{deming2013,wakeford2015,sing2016} and in particular the planetary mass-metallicity relation, a crucial piece of information in distinguishing among planet formation scenarios \citep[e.g.][]{kreidberg2014,thorngren2016,wakeford2017pancet}.  

Transmission spectroscopy is affected by condensates both in the visible and in the infrared. The WFC3/G141 spectral band ($1.1-1.7 \, \mu$m) provides a number of examples of spectral features whose amplitude is muted with respect to models of clear atmospheres and solar composition \citep[e.g.][]{deming2013,kreidberg2014_nat,wakeford2015,sing2016}. One possible explanation for this is the intrinsic low water abundance of such atmospheres, with implications on their formation conditions \citep{seager2005,madhusudhan2014_2,madhusudhan2014}. However, \cite{sing2016} present a comparative study of ten hot Jupiter atmospheres in both the visible and the infrared and provide evidence that clouds are the most likely explanation for the dampening of spectral features.

Aerosol formation is a strong function of the temperature, for which the equilibrium temperature \teq\ is used as a proxy, of the gravity $g$ of a planet, and of atmospheric composition. The giant planets characterized so far sit in a vast region of this phase space. They vary widely in temperature structure, which is determined by the radiative energy balance, and pressure structure of the atmosphere, mainly determined by the planet's gravity. Despite this, the number of characterized exoplanets is still low, molecular abundances and pressure-temperature (P-T) profiles are in most cases poorly constrained, allowing only empirical trends to be tentatively identified \citep[e.g.][]{stevenson2016}. Moreover, the population of characterized exoplanets lacks analyses of planets sharing similar parameters, which would enable testing whether models capture all the essential physics needed to correctly describe their aerosols.

\begin{figure}
\epsscale{1.3}
\plotone{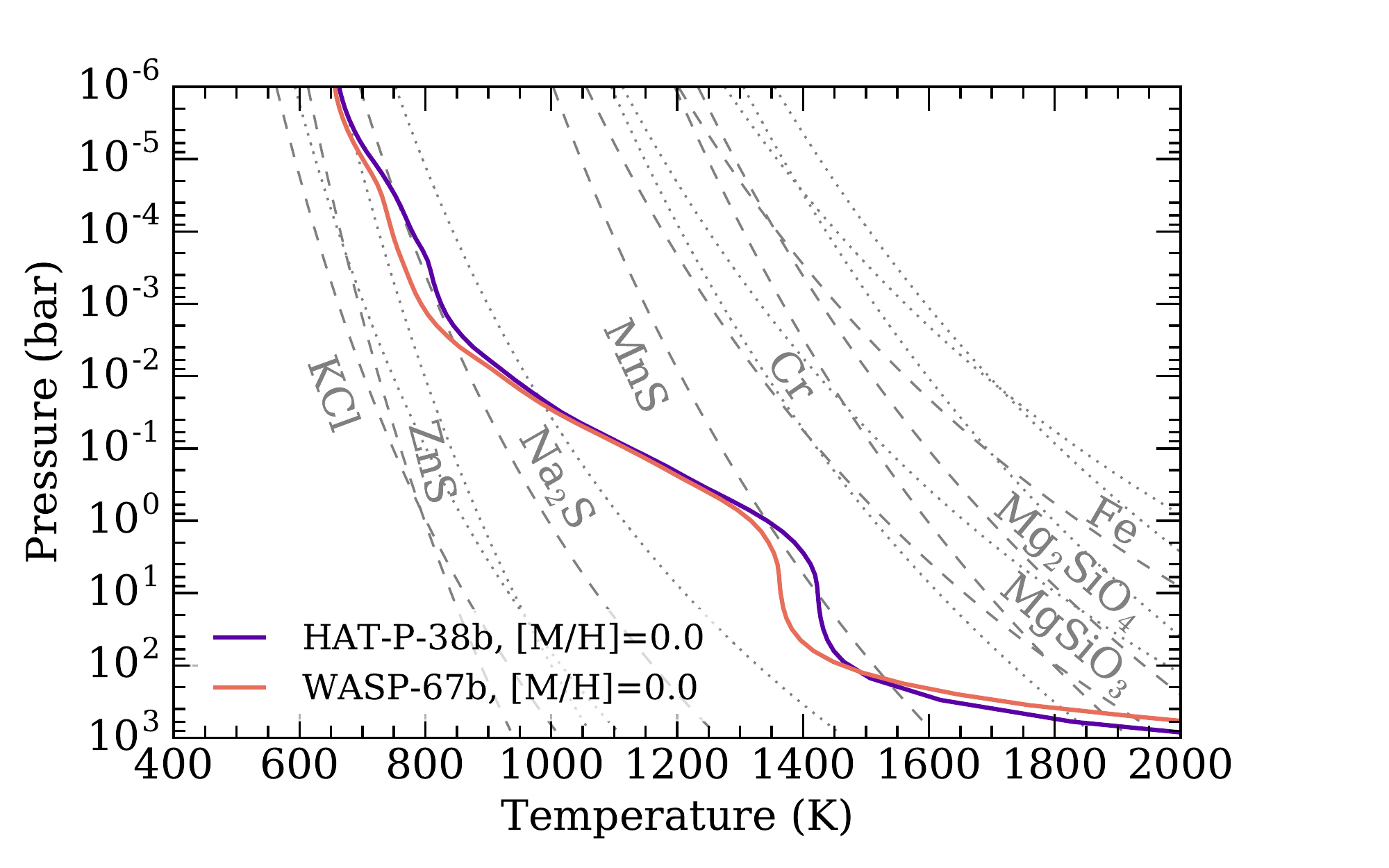}
\caption{P-T profiles of the two planets from this work and condensation curves from various molecules. From the intersection of the curves, alkali-bearing condensates are expected in the atmospheres of the planets. WFC3 observations probe the pressures betwenn about 1 and 100 mbars.}
\label{ptprofs}
\end{figure}

In the context of comparative planetology, the pair of giant planets WASP-67 b and HAT-P-38 b is an interesting benchmark. Orbiting similar stars (early K and late G, respectively) with very similar metallicities \citep{hellier2012, sato2012} at nearly equal orbital period (4.6 days), they share a similar level of insolation ($T_\mathrm{eq} \sim 1050$ K), unlike other planets in a similar region of the phase space. In addition, both planets are characterized by a gravity of about 10 $\mathrm{m \, s}^{-2}$. They are therefore expected to have the same bulk properties, as well as to exhibit nearly identical P-T profiles, which we calculate with 1D atmosphere models \citep{mckay1989,marley1996,fortney2008}, assuming solar composition and cloud-free atmospheres. From the profiles and the condensation curves shown in Figure \ref{ptprofs}, we expect the planets to be in a regime where alkali elements condense \citep{lodders1999,burrows2000,sudarsky2003,fortney2003,morley2012}. Comparatively explaining aerosol formation in these two giant planets, sharing an almost equal amount of insolation as well as gravity, is an essential step towards understanding H$_2$-He dominated atmospheres. Also, precise constraints on the molecular abundances of these planets would add a valuable contribution to our knowledge of the mass-metallicity relation and of planet formation.

In this work, we present a comparative analysis of the WFC3/G141 observations of WASP-67 b and HAT-P-38 b, collected during {\em {\em HST}} Cycle 23 (GO 14260, PI Deming).  We introduce the targets in closer detail in Section \ref{introtargets} and Section \ref{observations} describes the observations. We present the fits of the observations to derive the planets' transmission spectra in Section \ref{transitmodeling}, while in Section \ref{spectranalysis} we interpret the spectra though retrieval exercises. We discuss and conclude in Section \ref{future} and \ref{conclusions}, respectively.

\section{Targets}\label{introtargets}
WASP-67 b \citep{hellier2012} orbits a 10.6 mag$_J$, 0.87 $R_\odot$ K star ($T_\mathrm{eff} = 5200 \pm 100, \, \log g = 4.35 \pm 0.15, \, [\mathrm{Fe/H}] = -0.07 \pm 0.09$) with a 4.6-day orbital period. It is a 0.42 M$_\mathrm{J}$, 1.4 $R_\mathrm{J}$ inflated hot Jupiter with a grazing orbit (impact parameter $b > 0.9$). It is the first transiting exoplanet definitely known to have this characteristic. The grazing nature of its transit was confirmed by \cite{mancini2014}. This means the second and the third contact points are missing and causes the light curve solution to be degenerate. \cite{mancini2014} carried out a multi-band follow-up of this system and found a flat spectrum within the experimental uncertainties, in agreement with the predictions of \cite{fortney2008} for a planet of this temperature.

HAT-P-38 b \citep{sato2012}, with mass 0.27 M$_\mathrm{J}$ and radius 0.83 $R_\mathrm{J}$, orbits a 11.0 mag$_J$, 0.92 $R_\odot$, late G star ($T_\mathrm{eff} = 5330 \pm 100, \, \log g = 4.45 \pm 0.08, \, [\mathrm{Fe/H}] = 0.06 \pm 0.10$) with a period of 4.6 days. It is part of a rapidly increasing population of transiting Saturn-like planets with measured masses. At the time of writing, the \texttt{exoplanet.eu} archive contains 34 planets with $0.1 < M/M_J < 0.4$ and $0.7 < R/R_J < 1.1$, while only 14 of them were known at the time HAT-P-38 b was announced. The observed radius of this planet is slightly smaller than expected from the empirical relation of \cite{enoch2011}, who found a positive correlation between the radius of an exoplanet and its equilibrium temperature and a negative correlation between the radius and the metallicity of its host star. Following \cite{fortney2007}, \cite{sato2012} suggest its peculiarity could be related to the heavy-element composition of its core. However, other members of the same family of exoplanets contradict such trend, such as HAT-P-18 b and HAT-P-19 b \citep{hartman2011} or Kepler-16AB b \citep{doyle2011}.

\section{Observations and data reduction}\label{observations}


The WFC3 spectra of WASP-67 and HAT-P-38 are publicly available on the Mikulski Archive for Space Telescopes. \footnote{\texttt{https://archive.stsci.edu}} One visit of WASP-67 b was obtained on October 22, 2016 (transit duration of 1.9 hours, requiring four {\em HST} orbits per transit), and two visits of HAT-P-38 b were secured on March 2, 2016 and on August 26, 2016 (transit duration of 3 hours, requiring five orbits per transit). The WASP-67 b data can be obtained at \dataset[10.17909/T93D46]{https://doi.org/10.17909/T93D46} while the HAT-P-38 b data are at \dataset[10.17909/T9768W]{https://doi.org/10.17909/T9768W}.

A direct image of the targets was obtained with the F139M filter at the beginning of each {\em HST} orbit. The spectra were acquired with the G141 grating, operating between 1.075 and 1.700 $\mu$m with a resolving power of 130 at 1.4 $\mu$m. Seventeen spectra per {\em HST} orbit were acquired for WASP-67 and thirteen per orbit for HAT-P-38. The spectra were acquired in spatial scanning mode \citep{mccullough2012}, consisting in continuously nodding the telescope during the exposure. This allows for longer exposure times, therefore increasing the collected photons for bright targets and at the same time reducing overheads. Both targets were acquired in forward scanning direction only, with a scan rate of 0.037 arcsec s$^{-1}$ (0.28 pixel s$^{-1}$) for WASP-67 b and of 0.026 arcsec s$^{-1}$ (0.20 pixel s$^{-1}$) for HAT-P-38 b. We use the frames in IMA format, where non-destructive individual exposures, recorded every 22.3 seconds, are saved in separate \texttt{fits} extensions. Six non-destructive reads were obtained for each WASP-67 exposure, for a total exposure time of about 90 s, and eight for each HAT-P-38 exposure, for a total exposure time of about 134 s. The $256 \times 256$ pixels aperture was used in SPARS25 mode, yielding $\sim 34,000$ average electron counts per exposure for both targets. A first calibration of the raw images and correction for instrumental effects is carried out by the \texttt{calwf3} pipeline.

\begin{figure}
\epsscale{1.3}
\plotone{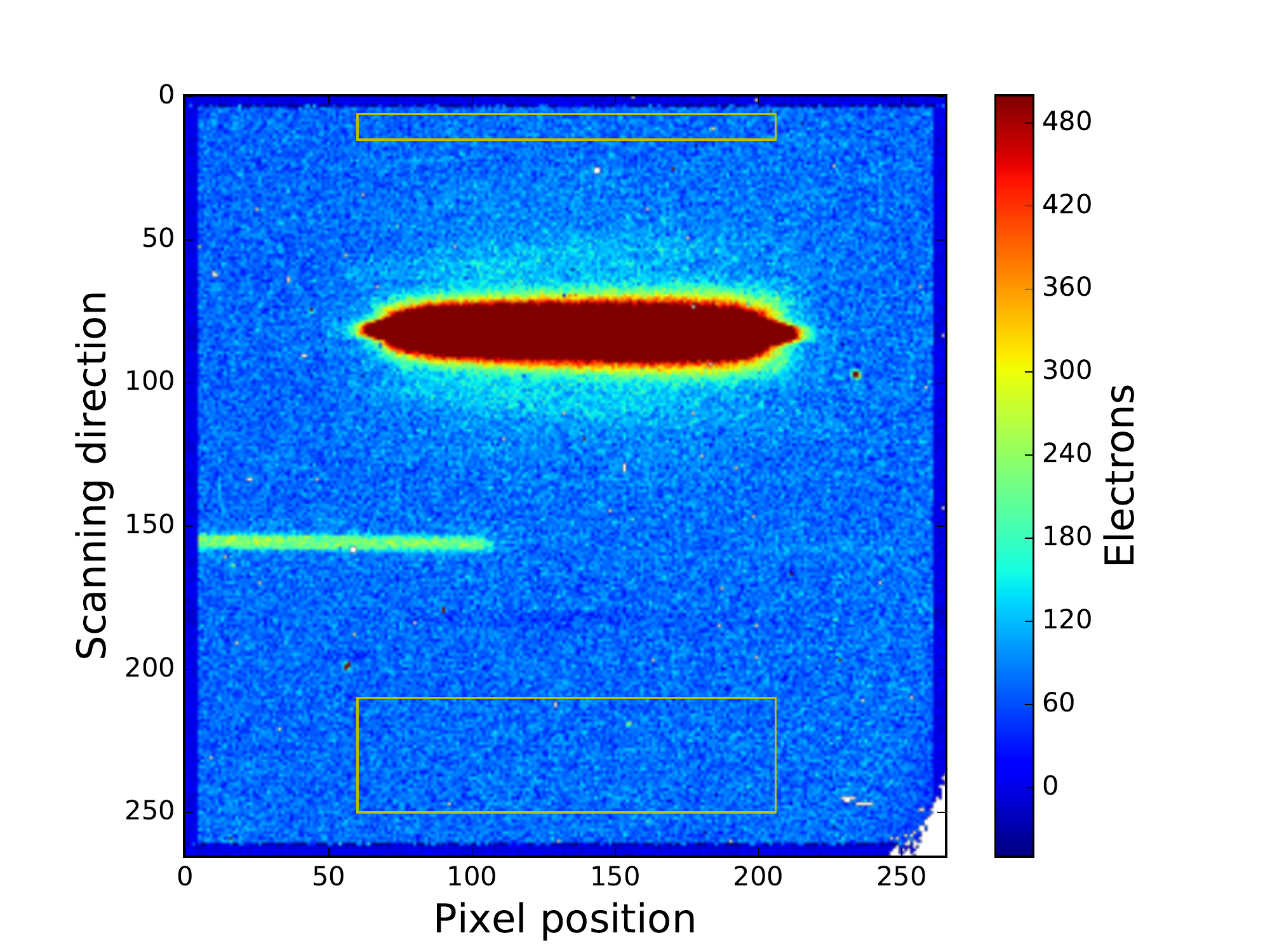}
\caption{Background flux in a frame taken during the first visit of HAT-P-38 b. The $x$-axis shows the values of column pixel, prior to wavelength calibration. The yellow boxes indicate the regions used for background estimation. A contaminant source is evident in the left part of the frame, as well as a residual charge in the central columns, which we avoided in the estimation of the background value. Plotted measurements are limited in the range indicated in the right column.}
\label{bkg1}
\end{figure}

\begin{figure}
\epsscale{1.2}
\plotone{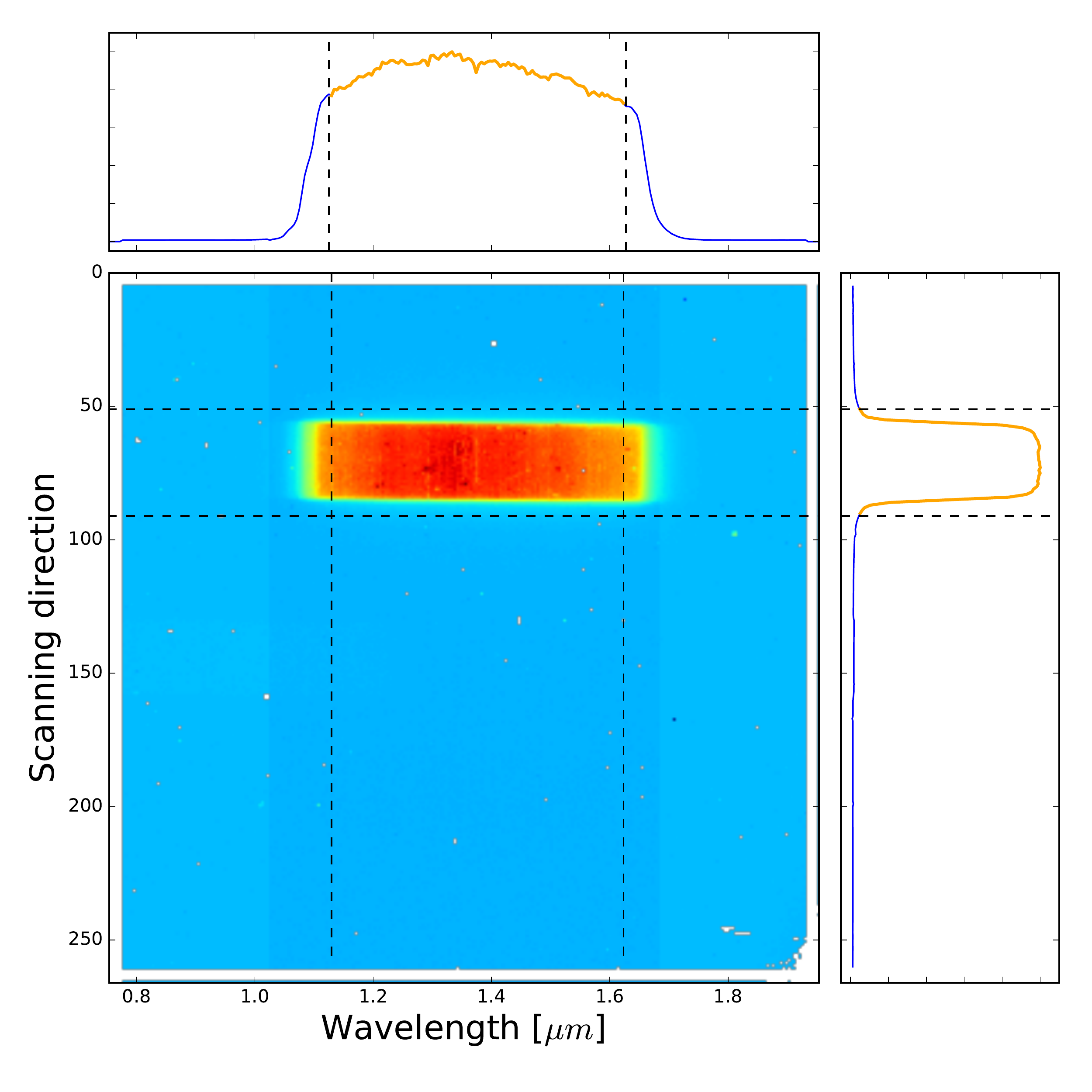}
\caption{The 2D spectrum presented in Figure \ref{bkg1} after background subtraction and optimal extraction. On the top plot, the 1D spectrum after integration over the frame columns. On the right, the spectral trace in the cross-dispersion direction. The dashed lines and the orange color show the limits chosen for wavelength and aperture integration. The flux range goes from 0 to about 38000 electrons.}
\label{reducedsp}
\end{figure}

The wavelength calibration is derived by using the centroid of the direct image of the first orbit and by applying the relations in \citet{pirzkal2016}. We differentiate, mask, and add consecutive non-destructive reads (NDR) following \cite{deming2013}. The background subtraction is performed column-by-column, by selecting regions of the detector which are both above and below the stellar spectrum. A column-by-column $5 \sigma$ clipping to reject cosmic rays and bad pixels is performed prior to the calculation of the column background median value. Figure \ref{bkg1} shows an example of background extraction for the first visit of HAT-P-38 b. The yellow boxes highlight the regions used for measuring the value of the background. For this visit in particular, we notice the presence of a contaminant source on the detector, which is clearly evident in the left part of the Figure. We also observe a residual charge in the central portion of the detector, which is less evident in the Figure. Our background correction and following extraction does not use the pixels affected by these effects. Each NDR is inspected for drifts both in the wavelength and in the scanning direction with the Python \texttt{image registration} package \citep{baker2001}. Then, each NDR is integrated using the optimal extraction algorithm \citep{horne1986} over an aperture determined by minimizing the residuals of the final white light curve from a transit model, as described in Section \ref{whitetr}.\\
We obtain a set of one-dimensional spectra (one per NDR) which are added after correcting a second time for drifts in the wavelength direction which were not corrected for in the previous phase, using a Fast Fourier Transform convolution algorithm to detect residual shifts. This is required because, even if \texttt{image registration} detects shifts in both directions of an image at the same time, the float precision it adopts can leave $\lesssim 0.1$ pixels shifts uncorrected. Figure \ref{reducedsp} shows the spectrum after the extraction of one of the IMA files.

The integrated flux between 1.125 and 1.650 $\mu$m, in orange in the top panel of Figure \ref{reducedsp}, yields the photometric flux measurement corresponding to each point of the band-integrated, or ``white'', light curves. Spectroscopic light curves are obtained by integrating the stellar spectra in channels, four to nine pixels (18.6 to 41.9 nm) wide. Repeating the transmission spectra extraction (Section \ref{spectrotr}) for various bin sizes allows us to assess the robustness of the reduction. While the bluemost wavelength used for the integration is kept fixed, different binnings produce $\sim10$ nm different wavelengths for the redmost pixel. A slightly different band-integrated transit is therefore obtained for each binning. By minimizing the reduced $\chi^2$ in the fit of such transits (section \ref{whitetr}), we finally adopt a binning of seven pixels for WASP-67 b and of six pixels for HAT-P-38 b. However, the final interpretation is not affected by the choice of the bin size.

\subsection{Ground-based observations}
Additional transit observations were obtained by a group of advanced amateur astronomers that helped us constrain the orbital parameters of the systems. A full transit of WASP-67 b was observed using the Manner-Vanderbilt Ritchey-Chr\'{e}tien (MVRC) telescope located at the Mt. Lemmon summit of Steward Observatory, AZ, on July 08, 2016 in the $r'$ filter. The observations employed a 0.6 m f/8 RC Optical Systems Ritchey-Chr\'{e}tien telescope and SBIG STX-16803 CCD with a 4k $\times$ 4k array of 9 $\mu$m pixels, yielding a $26\farcm6 \times 26\farcm6$ field of view and $0\farcs39$ pixel$^{-1}$ image scale. The telescope was heavily defocused, resulting in a typical ``donut'' shaped stellar PSF with a FWHM of $\sim 25$ arcsec. The maximum elevation above the horizon of the observations was $\sim$ $35^\circ$. There were passing clouds at times, and data points corresponding to significant losses in atmospheric transparency have been removed from the data set.

The Garlitz Private Observatory (GPO) was used to observe a full transit of HAT-P-38 b on September 27, 2016. GPO is a private observatory run by Joseph Garlitz in Elgin, Oregon. The instrument consists of a 304.8 mm aperture Newtonian telescope with a focal length of 1505 mm, resulting in an f/4.9 focal ratio. The telescope uses an SBIG ST402me CCD camera resulting in a $15.5' \times 10.3'$ field of view and a $1.22''$ pixel scale. A clear filter was used with an exposure time of 60 seconds. 363 science images were taken at 60 seconds exposure each, with 21 images eliminated as outliers. FWHM was approximately 6.4 arcseconds. The differential photometry settings used were an aperture radius of 10 pixels, 18 pixels for the inner annulus radius, and 27 pixels for the outer annulus radius. Sky conditions were clear, temperature was 283 K, the moon was 8\% waning, and image capture timing was synchronized using GPS.

We use the GPO observations of HAT-P-38 b to refine the ephemeris of the system, as discussed in the following Section.




\section{Transit modeling}\label{transitmodeling}

\subsection{White light transits}\label{whitetr}
Four {\em HST} orbits were required for the transit of WASP-67 b and five for each transit of HAT-P-38b. We follow the standard practice \citep[e.g.][]{deming2013} of discarding the first {\em HST} orbit from each transit, which is affected by considerably different systematics than the other ones. For the same reason, we also reject the first point of every orbit.\\
The band-integrated light curves are then modeled by combining a \cite{mandelagol2002} transit model, implemented in the \texttt{PyTransit} package \citep{parviainen2015}. The WFC3 systematics, i.e. planetary visit-long slope, {\em HST} breathing and ``ramp'' effect, are modeled with a combination of linear and exponential functions \citep[e.g.][]{berta2012,knutson2014,kreidberg2014,kreidberg2014_nat,fraine2014,stevenson2014_gj,wakeford2016,kilpatrick2017}. The fitted model is
\begin{align}
\begin{split}
    M(t) &= T(t, k_r, t_0, P, a/R_\ast, i, e, u_a, u_b) \\
         & \times (C + V \theta + B \phi) (1 - R \mathrm{e}^{-\psi/\tau}) \\
         & \times F_b
\end{split}
\label{transitsyst}
\end{align}

where $t$ is the time of the observation, $k_r$ the planet-to-stellar radius ratio, $t_0$ the transit midpoint, $P$ the orbital period, $a/R_\star$ the orbital semi-major axis normalized to the host star radius, $i$ the orbital inclination, $e$ the orbital eccentricity, $u_a$ and $u_b$ the linear and quadratic limb darkening coefficients. Because of the limited number of data points on the edges of the transits, we conservatively adopt a quadratic limb darkening law \citep{claret2000}: for our data sets, this does not produce appreciable differences in the fitted transit parameters with respect to more sophisticated limb darkening laws. 

The parameter $V$ accounts for a linear planetary visit-long trend, $B$ for an {\em HST} orbit-long trend, and $C$ is a vertical offset applied to the whole light curve. The ramp amplitude is modeled by the parameter $R$ and its timescale by $\tau$. The parameter $F_b$ represents the baseline flux. The parameter $\phi$ is the {\em HST} orbital phase and $\theta$ is the planetary phase. The parameter $\psi$ is instead the phase for the ramp feature, given by $2 \pi [ (t + t') \mod P_\mathrm{HST} ]/P_\mathrm{HST}$. The initial shift $t'$ tracks the state of the ramp at the beginning of the observations. It is found by trial and error and is fixed to 0.025 for WASP-67 b and to 0.03 for HAT-P-38 b during the fit of the other parameters.

The Differential Evolution Markov Chains algorithm 
of the \texttt{MC3} package \citep{cubillos2016} is used to sample the posterior distributions of the model parameters. An initial Levenberg–-Marquardt exploration is performed prior to the start of the chains and its result is used for the initial values of the jump parameters. We refer to the most updated papers \citep{hellier2012,sato2012} for the ephemeris and orbital parameters of each system $P, \, a/R_\ast, \, i$, and $e$, listed in table \ref{priortransits}. The parameters $a/R_\ast$ and $i$ are fixed, as the {\em HST} observations do not sample the transit profile edges. Their values are checked against a transit-only modeling performed with \texttt{PyTransit} and \texttt{MC3}, without systematics, of our MVRC and GPO observations. The corresponding transits and the respective best-fit models are shown in Figure \ref{transits_kelt}. We find agreement between the parameters $a/R_\ast$ and $i$ from the ground-based observations, indicated in Table \ref{priortransits}. Despite the smaller uncertainties in our fits, the published values are conservatively adopted.

Because of the possible biases in our current knowledge of the radius distribution of planets, we consider the planet-to-stellar radius ratio ($k_r$) as a scale parameter and use a Jeffrey prior as suggested by \cite{gregory2005}. We instead use a uniform prior for the transit midpoint ($t_0$) jump parameter. The distributions for both $k_r$ and $t_0$ are centered around the most recent values available from literature. For the purpose of the fit, the first point of each observation is set to 0 BJD. The limb darkening coefficients are linearly interpolated from PHOENIX specific intensity stellar models \citep{husser2013} for the $1.125-1.650 \, \mu$m wavelength range, using the mean values of $T_\mathrm{eff}, \, \log g,$ and $[\mathrm{Fe/H]}$ reported in Section \ref{introtargets}. Their values are listed in table \ref{limbdtable} and are fixed during the MCMC. Uniform priors are assigned to the systematics model parameters.

For each target, ten chains of $1 \times 10^6$ steps each are run for each visit, with $5 \times 10^4$ steps for the burn-in. The chains are thinned by a factor ten to reduce correlations among the parameters and their convergence is inspected with the Gelman-Rubin statistics \citep{gelman1992,brooks1998}. Because of the correlation among the exponenetial ramp amplitude and time scale, some chains converge to within 3\% of unity, instead of a desired 1\%. However, the best-likelihood parameters are achieved in a few $10^5$ steps at most, for all parameters. In this phase, the two visits of HAT-P-38 b are modeled separately to examine their consistency. 

\begin{table*}
\begin{center}
\caption{Priors for the transit parameters in the transit fits. $\mathcal{U}(a, b)$ and $\mathcal{J}(a, b)$ denote respectively a uniform and a Jeffreys distribution between $a$ and $b$.}
\label{priortransits}
\begin{tabular}{l|l|l}
\hline
                                                & WASP-67 b                         &   HAT-P-38 b     \\
\hline
Radius ratio $k_r$                              & $\mathcal{J}(0.07, 0.15)$         &   $\mathcal{J}(0.07, 0.15)$   \\
Transit midpoint $t_0$ [BJD]$^{(a)}$            & $\mathcal{U}(2457683.40, 2457683.49)$ &   $\mathcal{U}(2457449.52, 2457449.58)$ \\
Orbital period $P$ [days]                       & 4.61442 (fixed)                   &   4.640382 (fixed) \\
Normalized semi-major axis $a/R_\star^{(a)}$    & 12.78 (fixed)                     &   12.18 (fixed)     \\
Orbital inclination $i$ [degrees]$^{(b)}$       & 85.8 (fixed)                      &   88.3 (fixed)     \\
Orbital eccentricity $e$                        & 0 (fixed)                         &   0 (fixed)        \\
Limb darkening coefficients $u_a$, $u_b$        & 0.216, 0.220 (fixed)              &   0.122, 0.319 (fixed)    \\
\hline
\end{tabular}
\end{center}
\begin{tablenotes}\footnotesize
    \item $^{(a)}$ Fit on MVRC transit for WASP-67 b: $a/R_\star = 13.5 \pm 0.4$; $i = (85.7 \pm 0.02)^\circ$. $^{(b)}$ Fit on GPO transit for HAT-P-38 b: $a/R_\star = 11.5 \pm 1.0$; $i = (88.6 \pm 0.04)^\circ$.
\end{tablenotes}
\end{table*}

\begin{figure}[tbh]
\epsscale{1.27}
\plotone{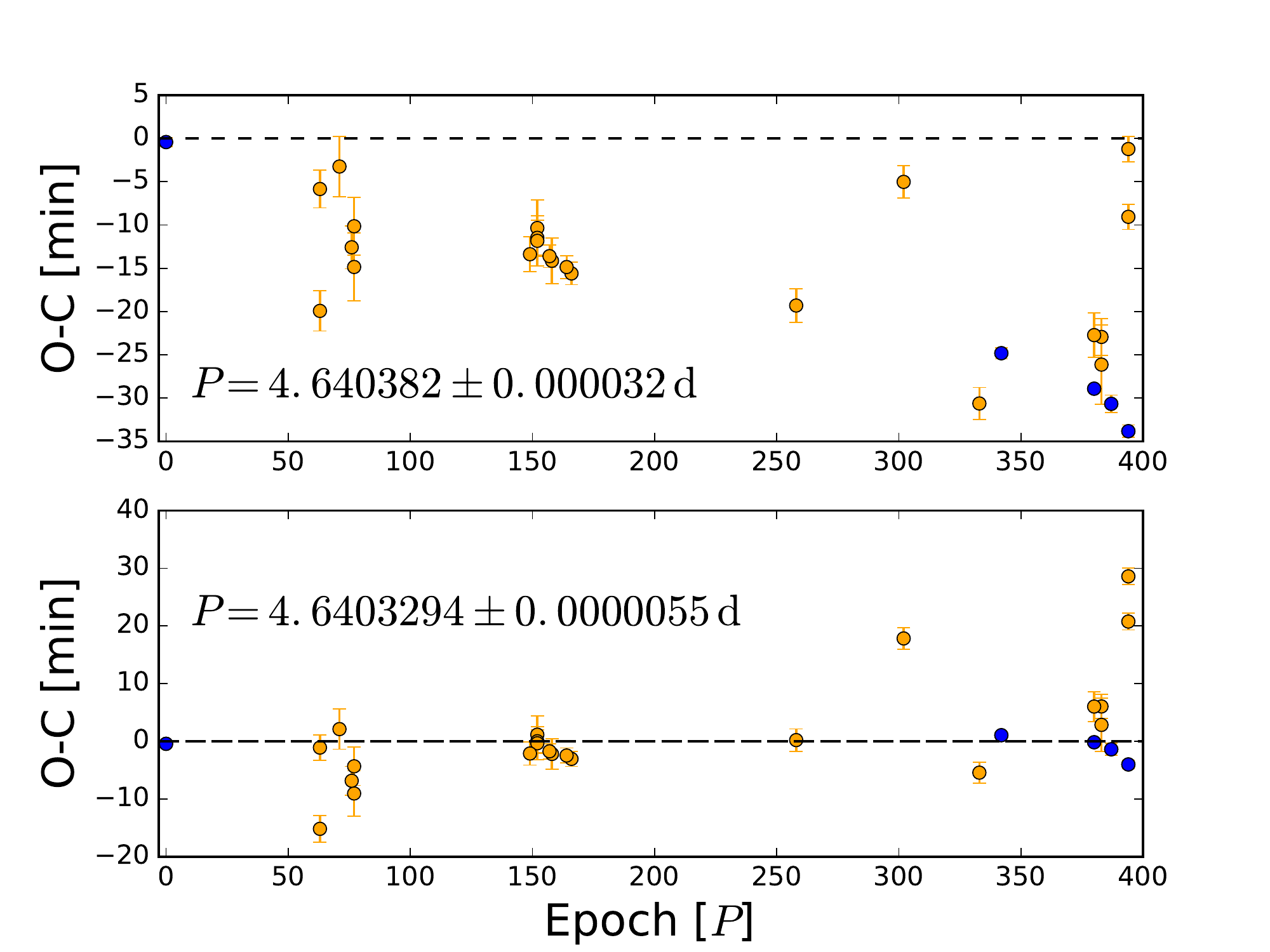}
\caption{$\mathrm{O-C}$ plot for HAT-P-38 b using the orbital period from \cite{sato2012} (top) and the one calculated with the transits in blue (bottom). The leftmost blue point in both panels corresponds to \citeauthor{sato2012}'s observation. The points in orange were not used for the calculation, due to their lower quality. Non-published values are extracted from the ETD.}
\label{omc}
\end{figure}

We compare the transit midpoints obtained from our observations, presented in Table \ref{whiteres} to the ground-based observations reported on the ETD database\footnote{\texttt{http://var2.astro.cz/ETD/index.php}}, as shown in Figure \ref{omc}. Our findings are consistent with a trend observed since 2011, which suggests the orbital period from \cite{sato2012} is slightly overestimated. We decide to refine the orbital period by using \citeauthor{sato2012}'s, our observations, and an observation collected on December 27, 2016 (BJD$_\mathrm{TDB} = 2457691.40735669 \pm 0.0005$) as part of the TRESCA program\footnote{\texttt{http://var2.astro.cz/EN/tresca/}} because of its good quality. The result, $P = 4.6403294 \pm 0.0000055 \, \mathrm{d}$, is $1.6\sigma$ lower than \citeauthor{sato2012}'s period and about six times more precise. Figure \ref{omc} shows that also the ground-based observations which were not used for the calculation, because of their lower quality (in orange) are consistent with this new period.

\begin{table*}
\begin{center}
\caption{Results of the MCMC on the white light curves.}
\label{whiteres}
\begin{tabular}{l|l|l|l}
\hline
                                                & WASP-67 b                         &   HAT-P-38 b - first visit  &  HAT-P-38 b - second visit   \\
\hline
Radius ratio $k_r$                              & $0.16066 \pm 2.1 \times 10^{-5}$  & $0.09285 \pm 1.9\times 10^{-4}$  & $0.09305 \pm 2.0\times10^{-4}$ \\
Transit midpoint $t_0$ [BJD (TDB)]                    & $2457683.98419 \pm 0.00013$       &  $2457450.11375 \pm 0.00045$ &  $2457626.44542 \pm 0.00010$ \\
Predicted $t_0$ from ground-based & & & \\
observations, using \cite{sato2012}'s  & & & \\
orbital period [BJD (TDB)]    &  $2457683.98224\pm0.00039$      &  $2457450.10969 \pm 0.00071$ &  $2457626.44420 \pm 0.00071$\\     
\hline
\end{tabular}
\end{center}
\end{table*}

\begin{figure}[tbh]
\epsscale{1.1}
\plotone{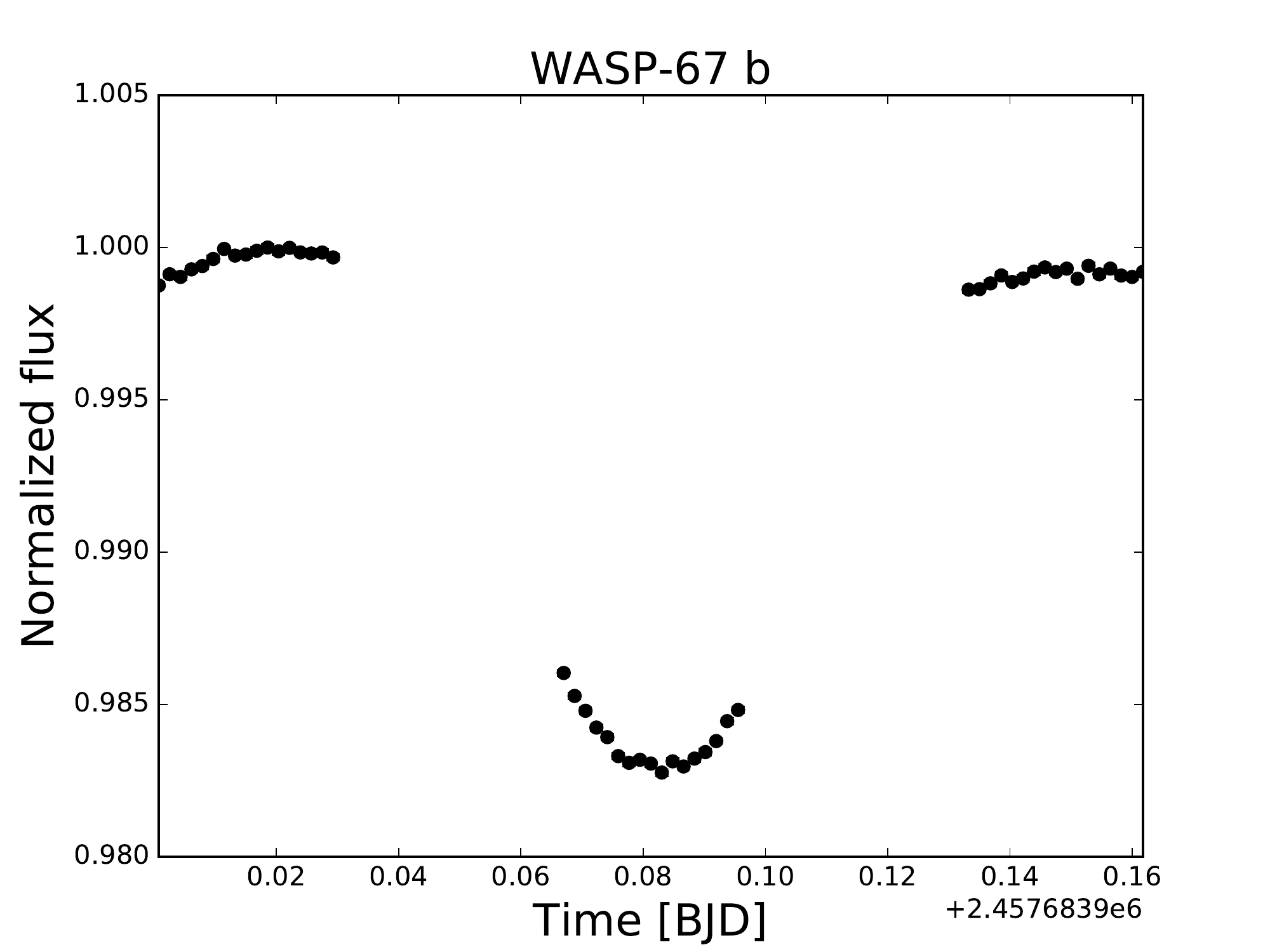}
\plotone{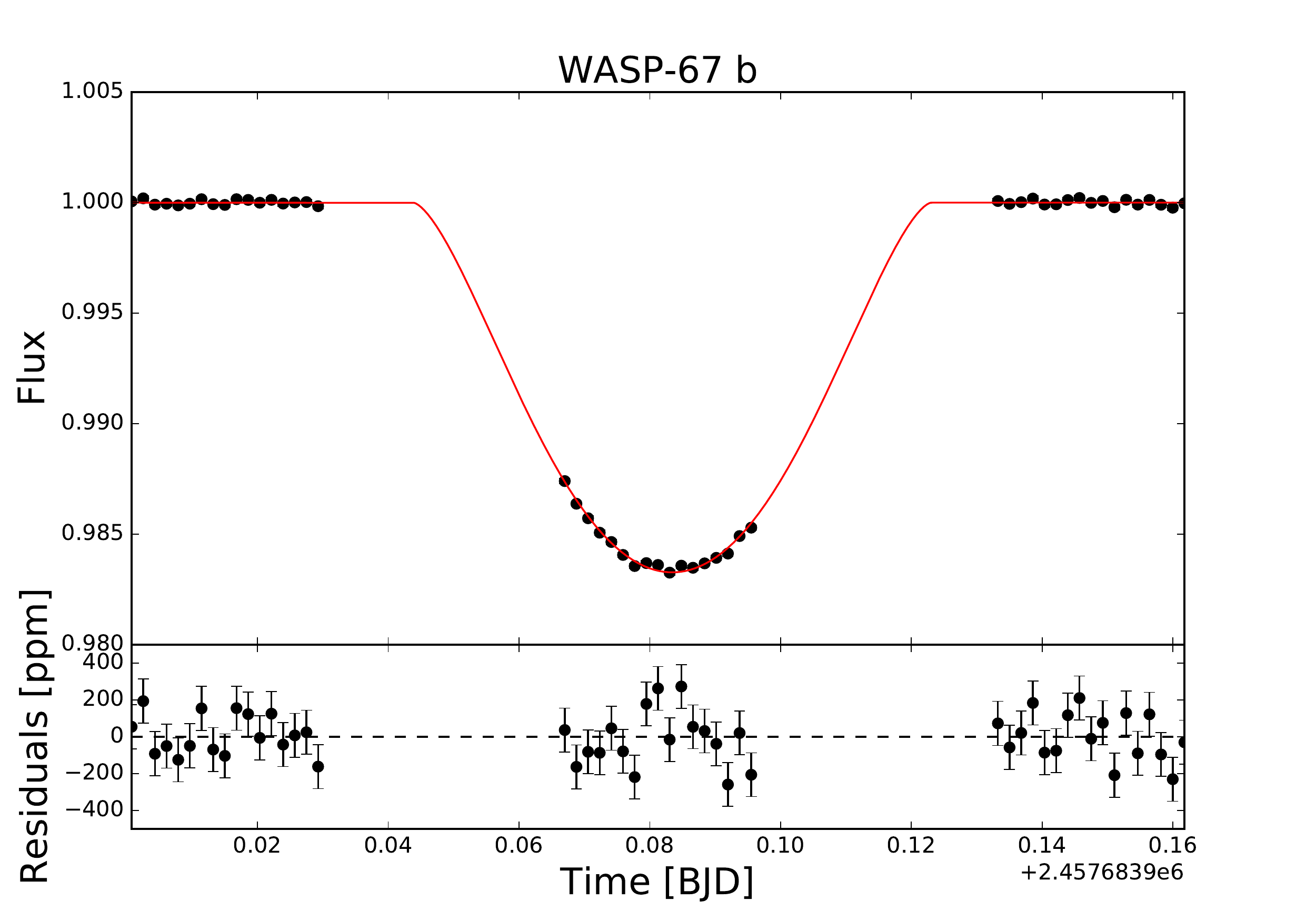}
\caption{$Top:$ Band-integrated transit of WASP-67 b prior to the correction for the systematics. $Bottom$: corrected white light curve (black points) and and best-fit model (red curve), after correction for the systematics. The residuals are shown in the lower panels.}
\label{w67white}
\end{figure}

\begin{figure}[tb]
\epsscale{1.1}
\plotone{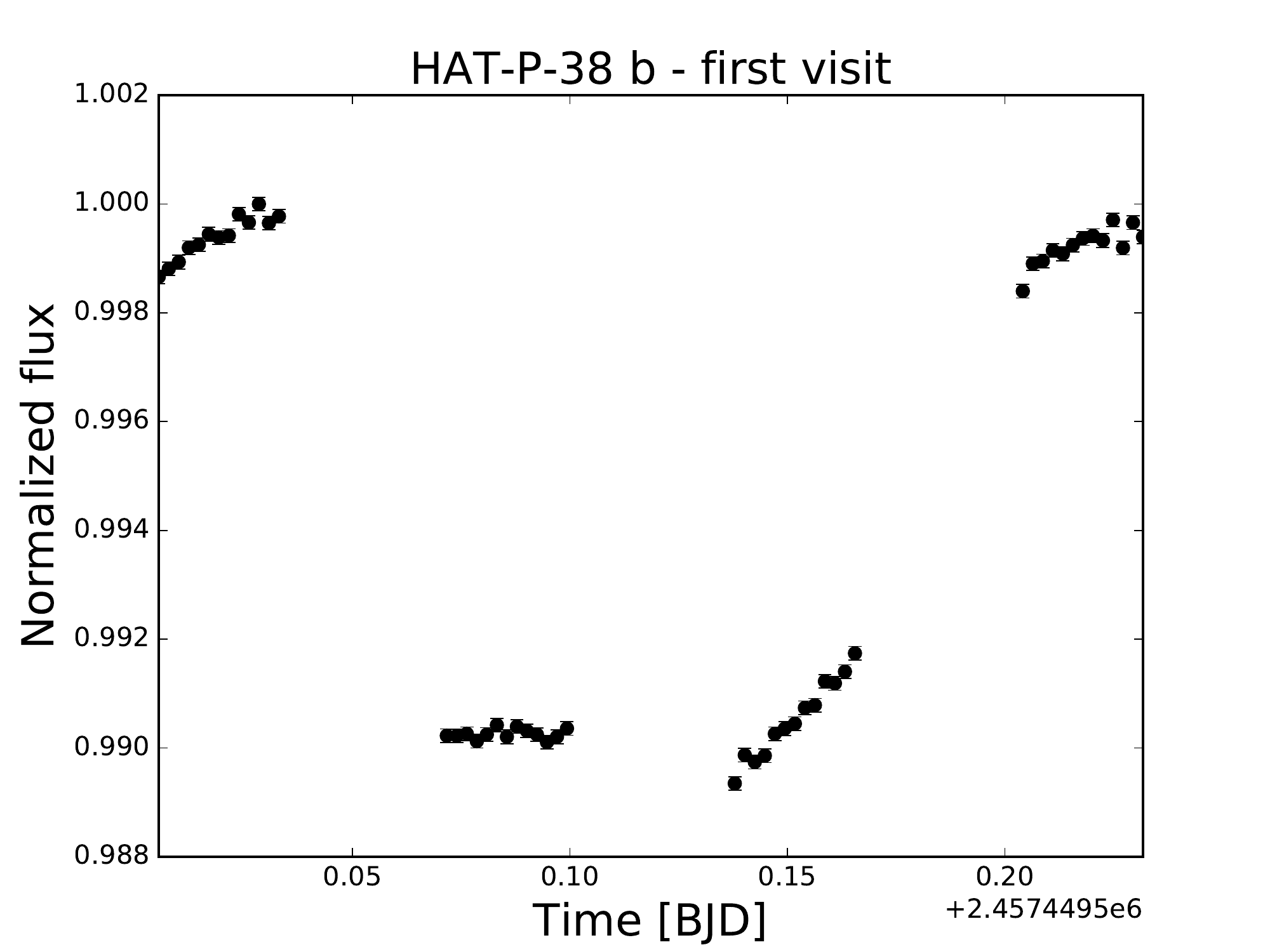}
\plotone{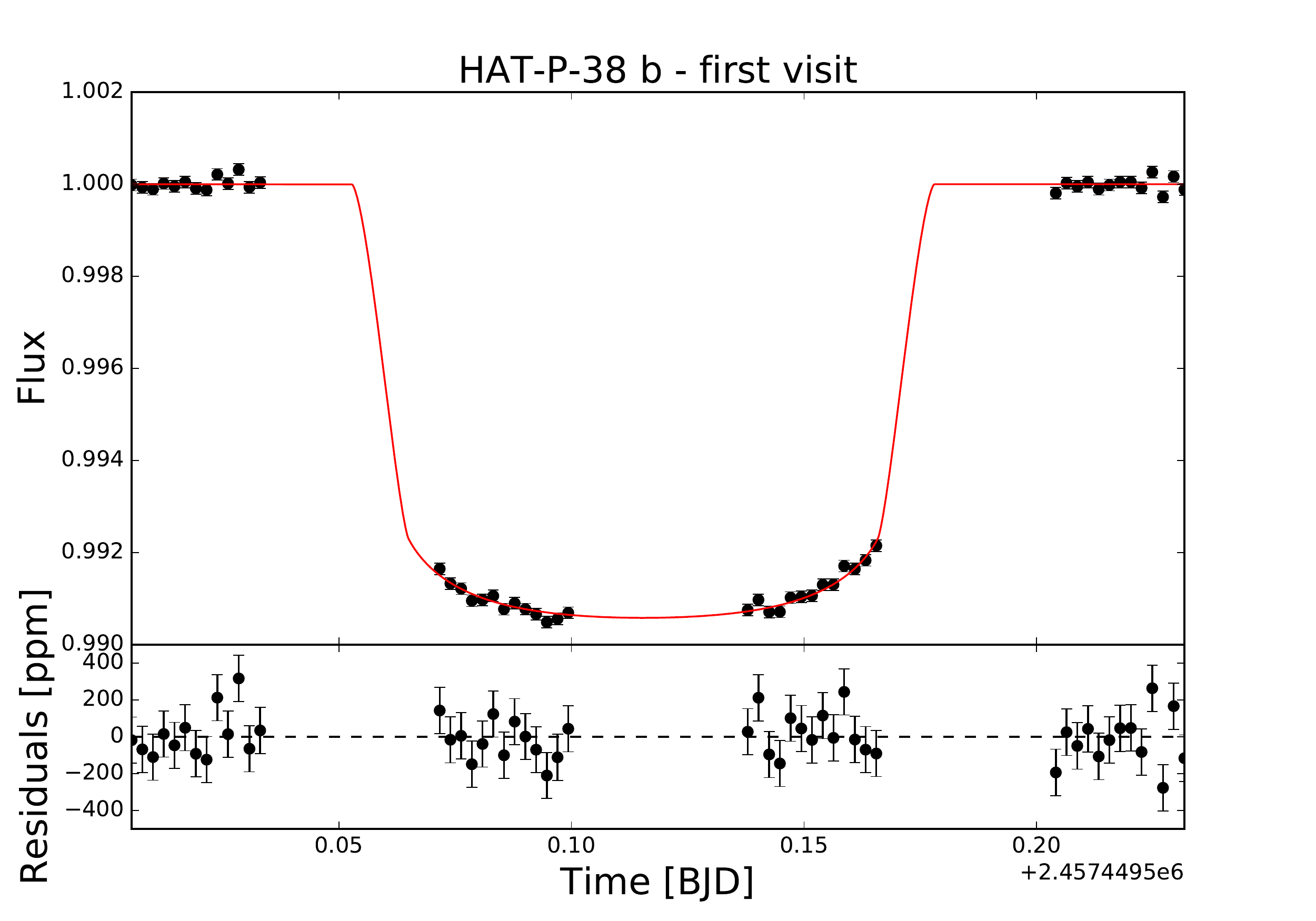}
\caption{As Figure \ref{w67white}, for the first visit of HAT-P-38 b.}
\label{h38white1}
\end{figure}

\begin{figure}[tb]
\epsscale{1.1}
\plotone{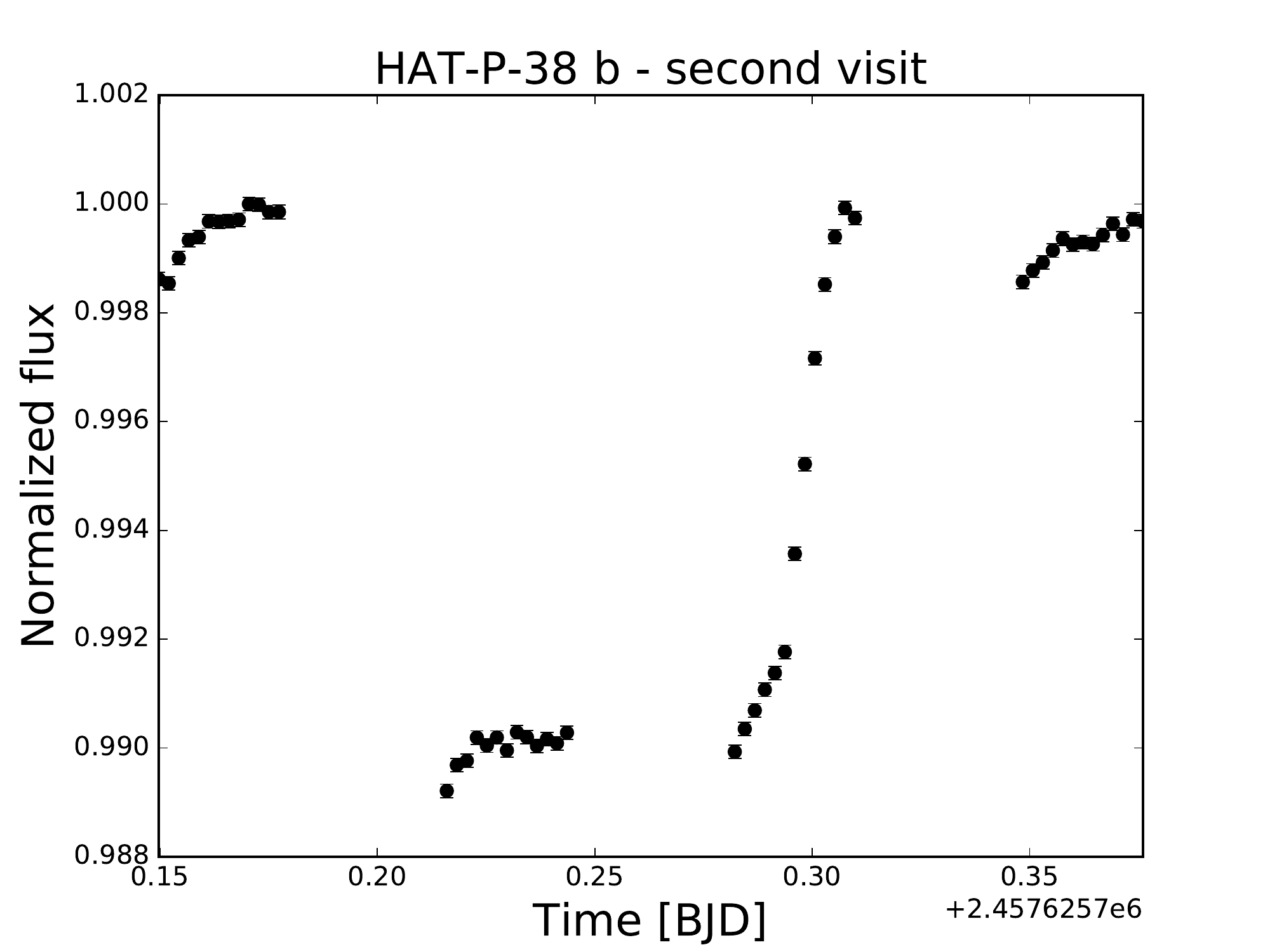}
\plotone{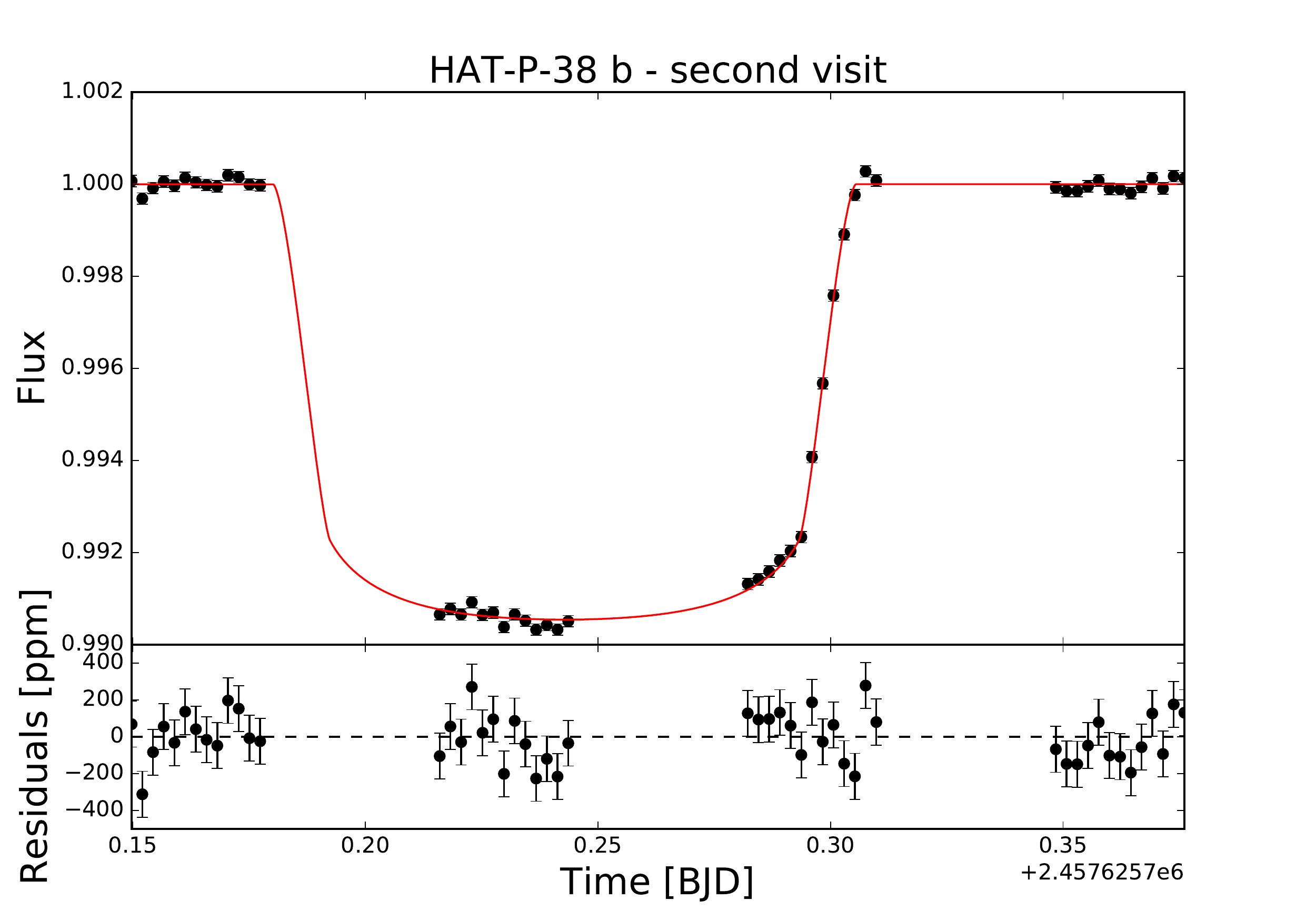}
\caption{As Figure \ref{w67white}, for the second visit of HAT-P-38 b..}
\label{h38white2}
\end{figure}

Figures \ref{w67white}, \ref{h38white1}, and \ref{h38white2} present the raw and corrected band-integrated transits with their respective best-fit models and residuals. We achieve a reduced chi-squared ($\tilde{\chi}^2$) of 1.39 for WASP-67 b. The values $\tilde{\chi}^2 = 1.10$ and $\tilde{\chi}^2 = 1.33$ are obtained for the first and second visit of HAT-P-38 b, respectively. The results are reported in Table \ref{whiteres}. 

\subsection{Spectroscopic transits}\label{spectrotr}
The spectroscopic light curves are obtained by integrating the stellar spectrum in channel widths of four to nine pixels (section \ref{observations}). 
For each one of such light curves we perform a common-mode correction of the systematics \citep{stevenson2014_w12}, that is, each transit is divided by the residuals of the best-likelihood model of the white light curve resulting from the MCMC. The parameter $k_r$ is set as a jump parameter and the transit midpoint $t_0$ is fixed to the band-integrated light curve best fit. A linear slope is used to model visit-long trends unaccounted for by the white light curve fit. The limb darkening coefficients are interpolated, as previously described, for each spectral channel, and fixed. Their values are reported in table \ref{limbdtable}.

\begin{figure}[htb]
\epsscale{1.3}
\plotone{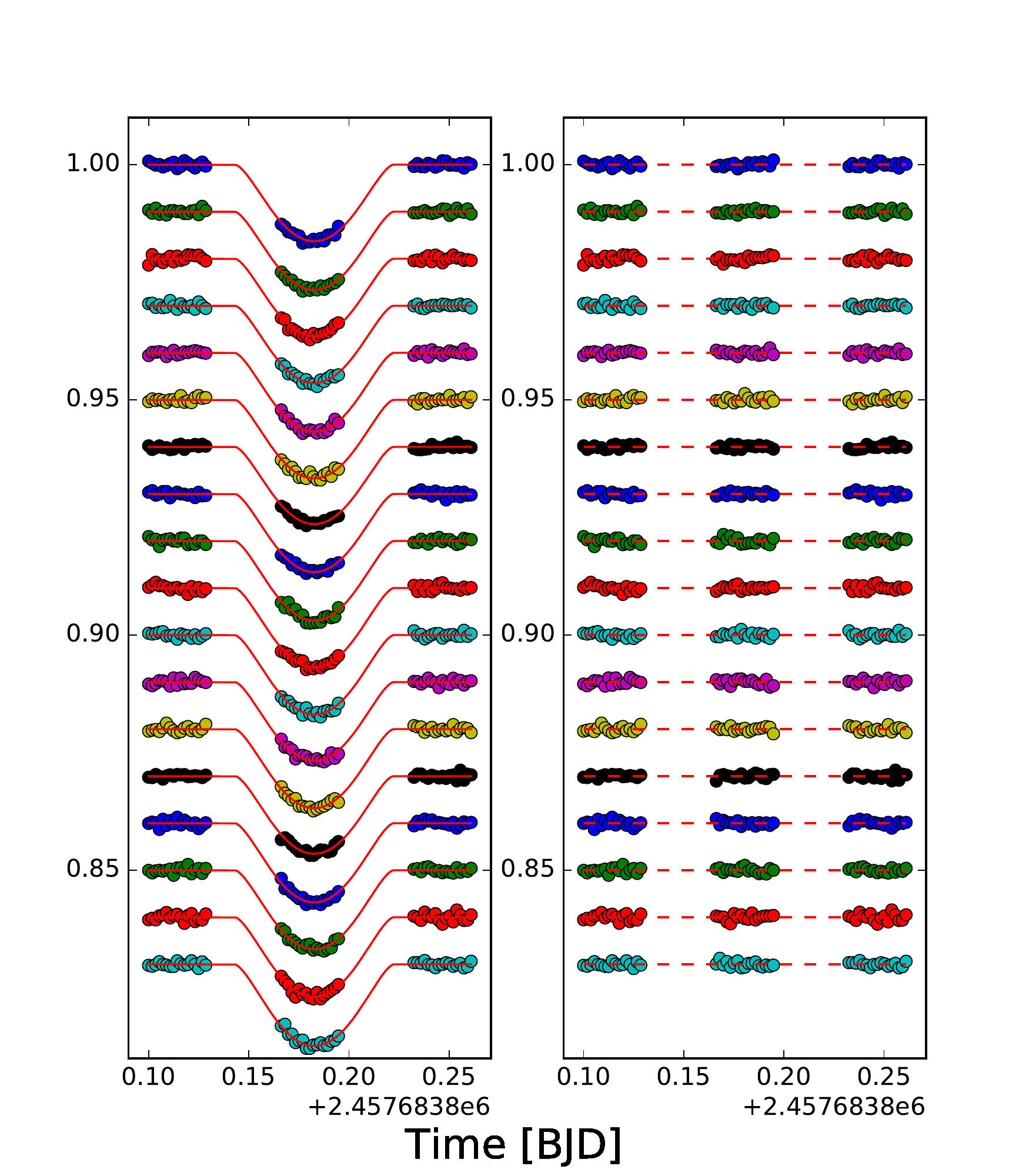}
\caption{Spectroscopic transits of WASP-67 b after correction for the systematics and best-fit models. Spectrophotometric channels are shifted for clarity. On the right panel, residuals for each transit, shifted to the continuum level of the respective transit.}
\label{transitsw67}
\end{figure}

\begin{figure}[htb]
\epsscale{1.3}
\plotone{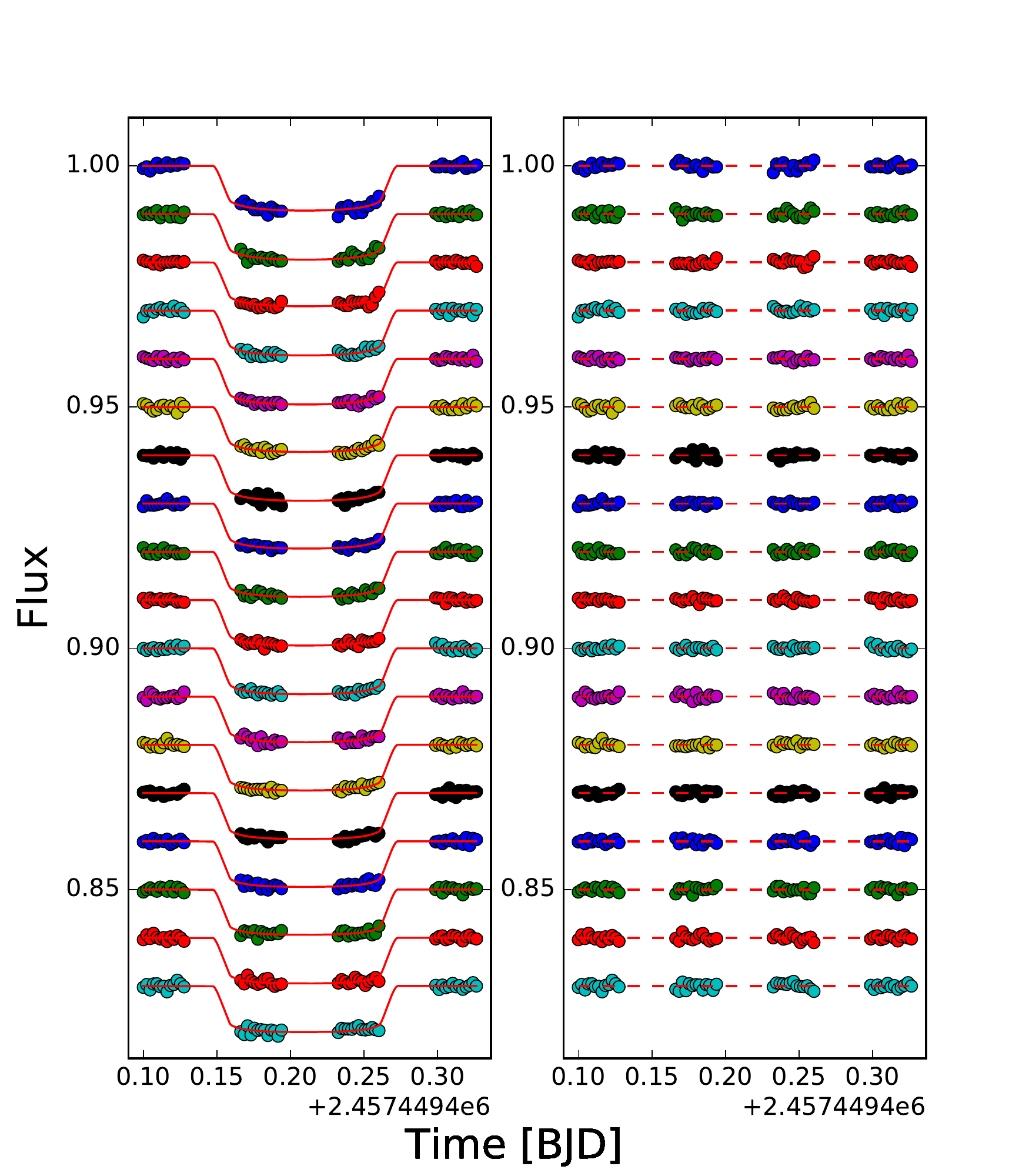}
\epsscale{1.3}
\plotone{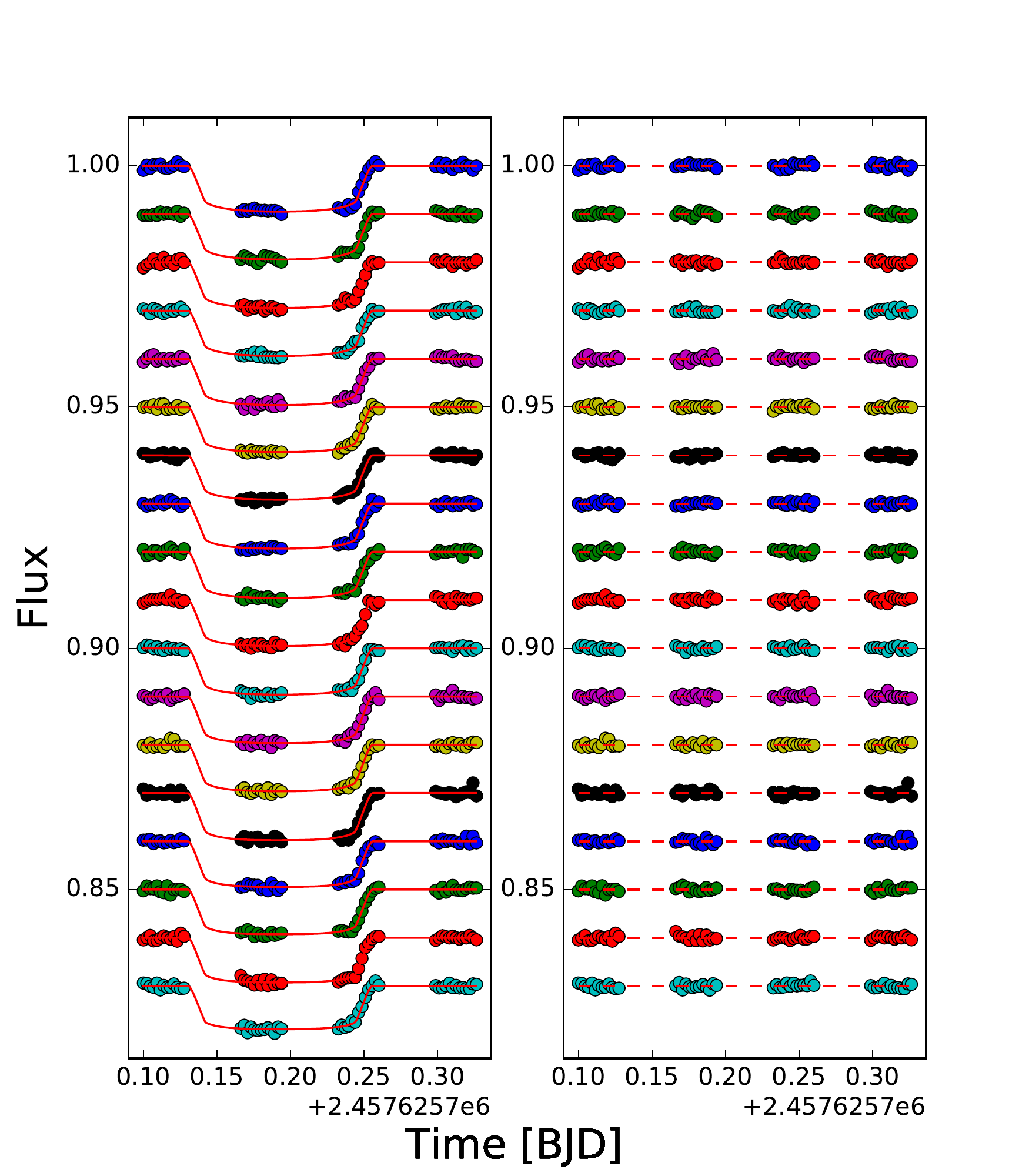}
\caption{As Figure \ref{transitsw67}, for HAT-P-38 b. First and second visit on top and bottom panel, respectively.}
\label{transitsh38}
\end{figure}

Figures \ref{transitsw67} and \ref{transitsh38} present the spectroscopic transits after correction for both the white light curve systematics and the baseline and linear function just described. The fitted transit depths are reported in Tables \ref{radiusratio_w67} and  \ref{radiusratio_h38}.\\

\begin{figure*}[tb]
\epsscale{1.2}
\plotone{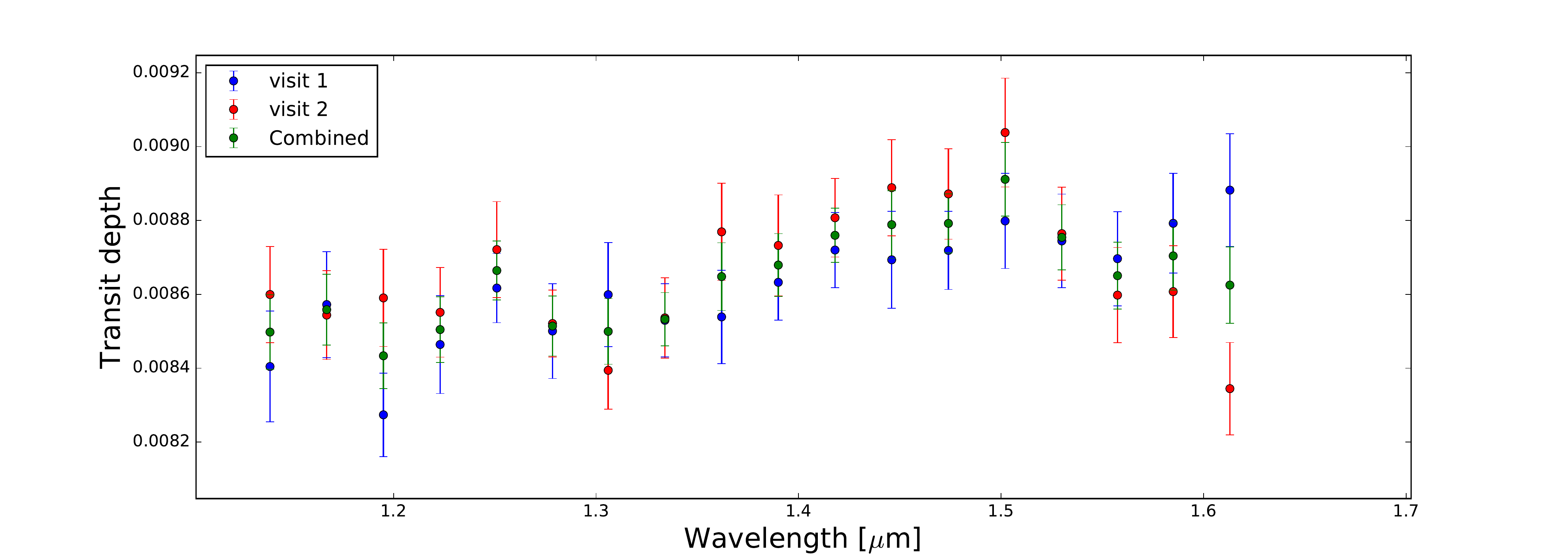}
\caption{Transmission spectra for the first and second visit of HAT-P-38 b, and from the joint modeling of the two visits.}
\label{visits_comp}
\end{figure*}

A homogeneous reduction of the two visits of HAT-P-38 b produces two slightly different transmission spectra. The first visit presents a slope in the spectrum of unidentified origin, which significantly mutes the water absorption feature with respect to the second visit. Despite this, almost all points of the two spectra, considered independently, agree at $1 \sigma$, as shown in Figure \ref{visits_comp}. The shift of the two visits is more important for $\lambda > 1.55 \, \mu\mathrm{m}$, as will be later discussed in Section \ref{twovisits}. The difference between the spectrum of visit 2 and visit 1, where the channels with $\lambda > 1.55 \, \mu\mathrm{m}$ are excluded, divided by the noise level calculated as the sum of the point-by-point variances, is on average $\simeq 0.6$, therefore not significant. This is consistent with the difference in $k_r$ of the two band-integrated transits, with the noise level calculated in the same way, i.e. $\simeq 0.7$.

The robustness of this difference is verified by varying the background region in the image reduction phase (Section \ref{observations}), the threshold for the rejection of cosmic rays and bad pixels on the detector, and by repeating the MCMC run with various $a/R_\star$ and $i$, compatibly with the uncertainties reported in literature. Different models for the systematics are also tested (e.g., second-order polynomials for the visit-long trend and polynomial functions; see \citealp{wakeford2016}), without solving the difference of the two spectra. Moreover, our residuals do not present indications of red noise that might affect the shape of the transmission spectra. The correlation matrices given by the residuals of the common-mode corrected spectroscopic transits and by their respective models are each compared to correlation matrices representing residuals affected only by white noise (that is, extracted from Gaussian distributions with zero mean and $\sigma$ equal to the standard deviation of the residuals in each channel). The distance between the correlation matrix of each visit and the white-noise correlation matrix is measured with the metric \citep{herdin2005}
\begin{equation}
    d_\mathrm{corr} (\boldsymbol{V}, \boldsymbol{G}) = 1 - \frac{\mathrm{tr} \left\{ \boldsymbol{V} \boldsymbol{G} \right\} }{||\boldsymbol{V}||_f ||\boldsymbol{G}||_f} \in [0, 1],
\end{equation}
where $\boldsymbol{V}$ and $\boldsymbol{G}$ are the correlation matrices for the residuals and the one representing white noise, respectively. For $f$, we use the Frobenius norm \citep{golub1996}. The distance is 0 for matrices equal up to a scaling factor, and 1 for a maximum extent difference. We measure $d_\mathrm{corr} < 0.3$ for both visits, with lower values when considering only parts of the matrices (that is, selecting only some channels), and never find any significant difference for the two visits.

We also highlight that the reduced $\chi^2$ of the transit fit on the spectroscopic channels does not provide indications of the presence of red noise. The reduced $\chi^2$ for the two visits are reported in table \ref{chisq}, which yield $\tilde{\chi^2} = 1.005 \pm 0.023$ for the first visit and $1.017 \pm 0.075$ for the second one.
We therefore choose to combine the two visits and measure a third spectrum by jointly modeling the two visits in each spectroscopic channel. In the respective MCMC, the radius ratio is shared among the two visits. The resulting combined spectrum is shown as well in Figure \ref{visits_comp} and the results of the MCMC are in Table \ref{radiusratio_h38}. 

\section{Transmission spectra analysis}\label{spectranalysis}

\subsection{Water absorption significance}
\begin{figure*}
\epsscale{0.55}
\plotone{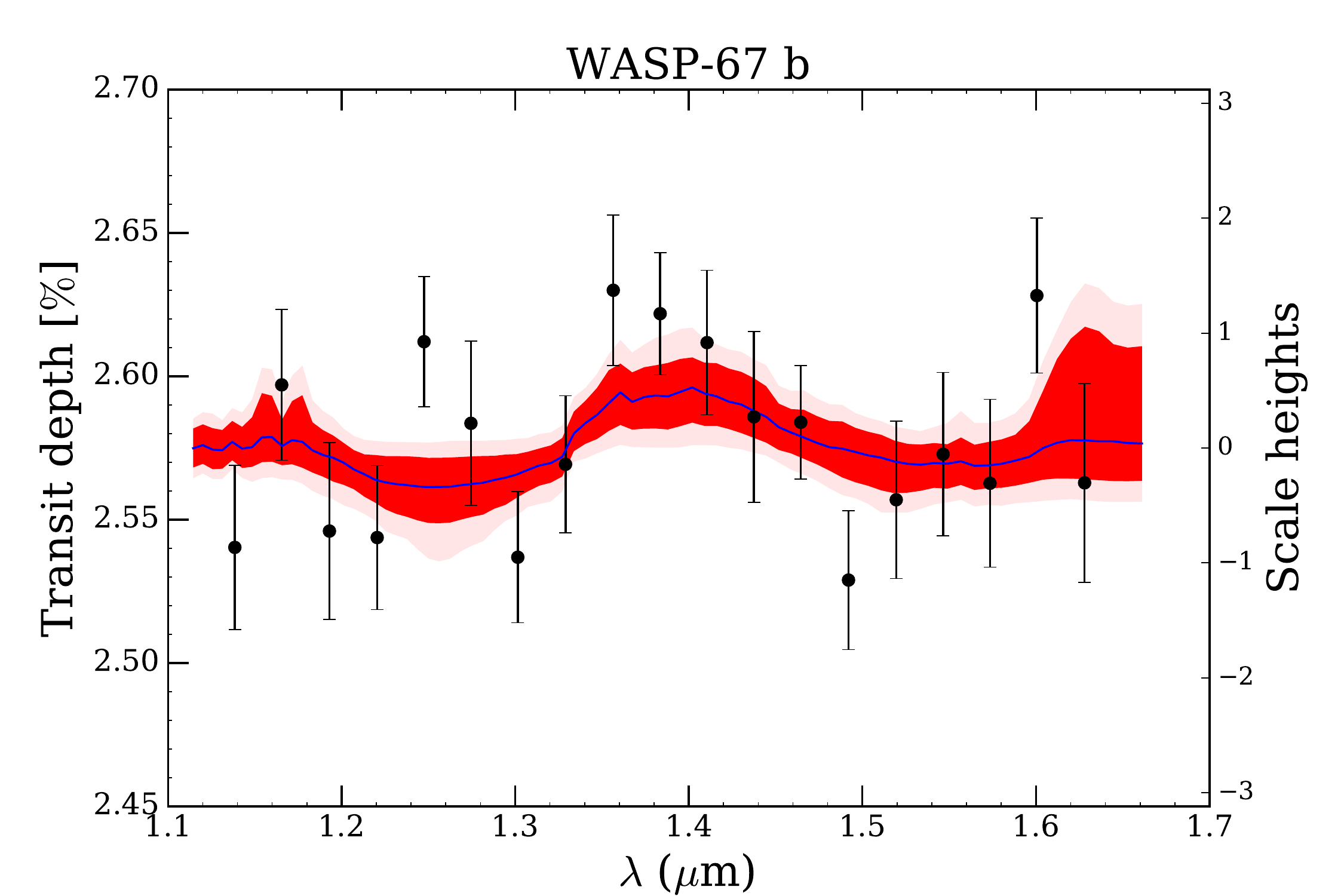}
\epsscale{0.55}
\plotone{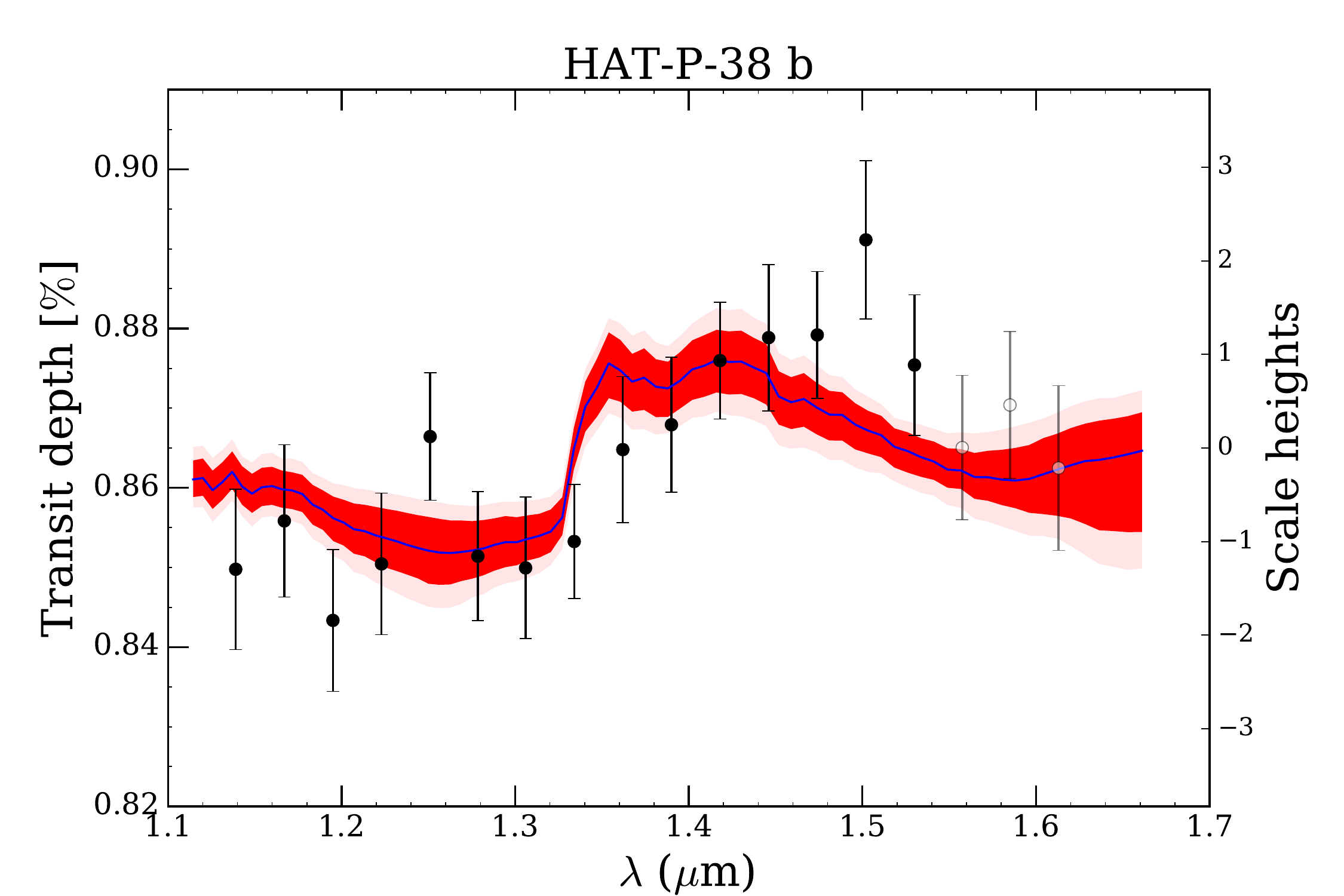}
\caption{Best-likelihood solution (blue), 1-- and $2\sigma$ (shades of red) confidence regions for the retrieved models on WASP-67 b data ($left$) and HAT-P-38 b data ($right$). The empty circles for HAT-P-38 b correspond to the points which are rejected for the retrievals (Section \ref{twovisits}). The atmospheric scale heights corresponding to the spectral features are on the right of each plot.}
\label{spectraconfidence}
\end{figure*}

In Figure \ref{spectraconfidence}, the G141 transmission spectra of WASP-67 b and HAT-P-38 b are compared (see Section \ref{sect_retrievals} for a discussion of the different markers and of the models plotted on the top of the spectra). The significance of the spectral features depends on the scale height of each atmosphere.

We use the values in \cite{hellier2012} for the mass, radius, equilibrium temperature ($1040 \pm 30$ K with zero Bond albedo), and gravity ($g \simeq 5.0 \, \mathrm{m \, s}^{-1}$) in order to derive the scale height $H$ for WASP-67 b. This yields $H = k_B T_\mathrm{eq}/(\mu g) \simeq 750 \, \mathrm{km}$, where $k_B$ is Boltzmann's constant and where we adopt a Jupiter-like mean molecular weight $\mu$ (2.3 u), One scale-height change in altitude corresponds to a change in transit depth of
\begin{equation}
    D \simeq k_r^2 \sim 2 k_r \frac{H}{R_\ast} \simeq 4 \cdot 10^{-4}.
\end{equation}
The water absorption amplitude is estimated with the \htoj\ index, following \cite{stevenson2016}. The average transit depth between 1.36 and 1.44 $\mu$m is used as the measure of the water absorption peak significance. For the J band, i.e. the baseline, we use the 1.22-1.24 $\mu$m wavelength range instead of the one used by \citeauthor{stevenson2016} (1.22-1.30 $\mu$m), because of the presence of an absorption feature in that channel. This yields an index of $1.5 \pm 0.5$, which indicates the likely presence of obscuring clouds in the atmosphere of WASP-67 b according to \citeauthor{stevenson2016}'s classification.\\
The same calculation is repeated for the combined visits of HAT-P-38 b, with $T_\mathrm{eq} = 1082 \pm 55$ K with zero Bond albedo and $g \simeq 10 \, \mathrm{m \, s}^{-1}$ \citep{sato2012}. We obtain $H \simeq 420$ km and $D \simeq 1.2 \cdot 10^{-4}$. This implies an \htoj\ index of $1.7 \pm 0.9$, suggesting a lower level of muting of spectral features due to condensates.

Given the uncertainties on the indexes, we refrain from giving any interpretation on their values, in particular with reference to the trends observed by \cite{stevenson2016}. The large error bars are due to the uncertainties on the measured transit depths and an additional level of uncertainty on the cloudiness of these atmospheres is given by the cloud top pressure--metallicity degeneracy \citep[e.g.][]{seager2000,charbonneau2002,fortney2005,benneke2012,sing2016}, which \citeauthor{stevenson2016}'s index cannot capture. We explore the degeneracy in more detail with retrieval exercises, described below.

\section{Retrievals}\label{sect_retrievals}
We use retrievals to explore the molecular abundances, atmospheric temperature, and cloud top pressure allowed for by observations \citep[e.g.][]{benneke2012,line2013,kreidberg2015}. Our retrievals are performed with the CHIMERA suite \citep{line2013}, which uses the nested Bayesian sampler \texttt{PyMultiNest} to derive the posterior distributions of the model parameters \citep{feroz2009,buchner2014}. In its formalism, a transmission spectrum is described by three parameters determining the P-T structure of the atmosphere (irradiation temperature $T_\mathrm{irr}$, IR opacity $\log \kappa_\mathrm{IR}$, ratio of visible to IR opacity $\log \gamma_1$), two determining global and relative abundances (metallicity $\log [M/H]$ and carbon-to-oxygen abundance ratio $\log (\mathrm{C/O})$), two accounting for disequilibrium processes through the vertical quench pressures for carbon and nitrogen ($\log P_\mathrm{q} (\mathrm{C})$ and $\log P_\mathrm{q} (\mathrm{N})$), two for the scattering cross section and slope ($\sigma_0$ and $\gamma$), and one for a gray cloud top pressure ($\log P_\mathrm{c}$). A scaling factor for the planetary radius at 10 bar ($xR_p$) is used to take into account the lack of absolute normalization for the modeled spectrum. 

As the available wavelength coverage offers little constraint on the thermal structure of the atmosphere, on the quench pressures, and on scattering, we adopt a nearly isothermal profile by fixing $\log (\kappa_\mathrm{IR})$ and $\log \gamma_1$. As the data encodes no information on quenching processes, we turn off quenching by setting the quench pressures to the $\mu$bar value. Finally, as scattering does not contribute to spectral features in the near IR, we impose no scattering by fixing a negligible scattering cross section and setting $\gamma = -4$. We adopt uniform priors for all the other parameters, as listed in table \ref{priorsretrievals}.

We perform at first a retrieval with the assumption of chemical equilibrium. Retrievals where this assumption is relaxed are presented, too. In these latter, metallicity and C/O ratio are fixed to their solar values ($[\mathrm{M/H}] = 0.0$ and $-0.26$, respectively), while water and methane abundances are free parameters and do not scale with the abundances of the other modeled molecules.

\subsection{The two visits of HAT-P-38 b}\label{twovisits}
\begin{figure}
\epsscale{1.0}
\plotone{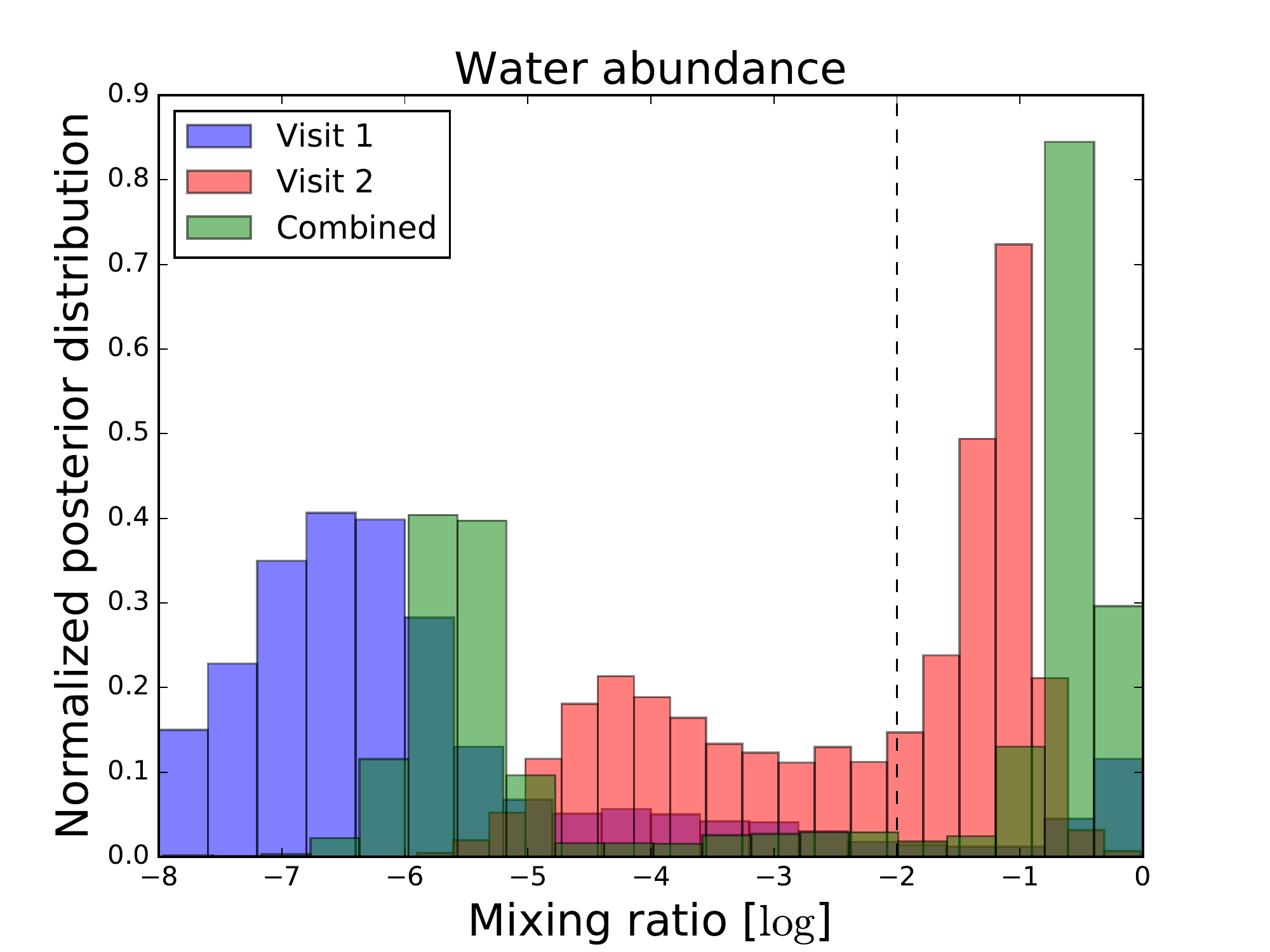}
\plotone{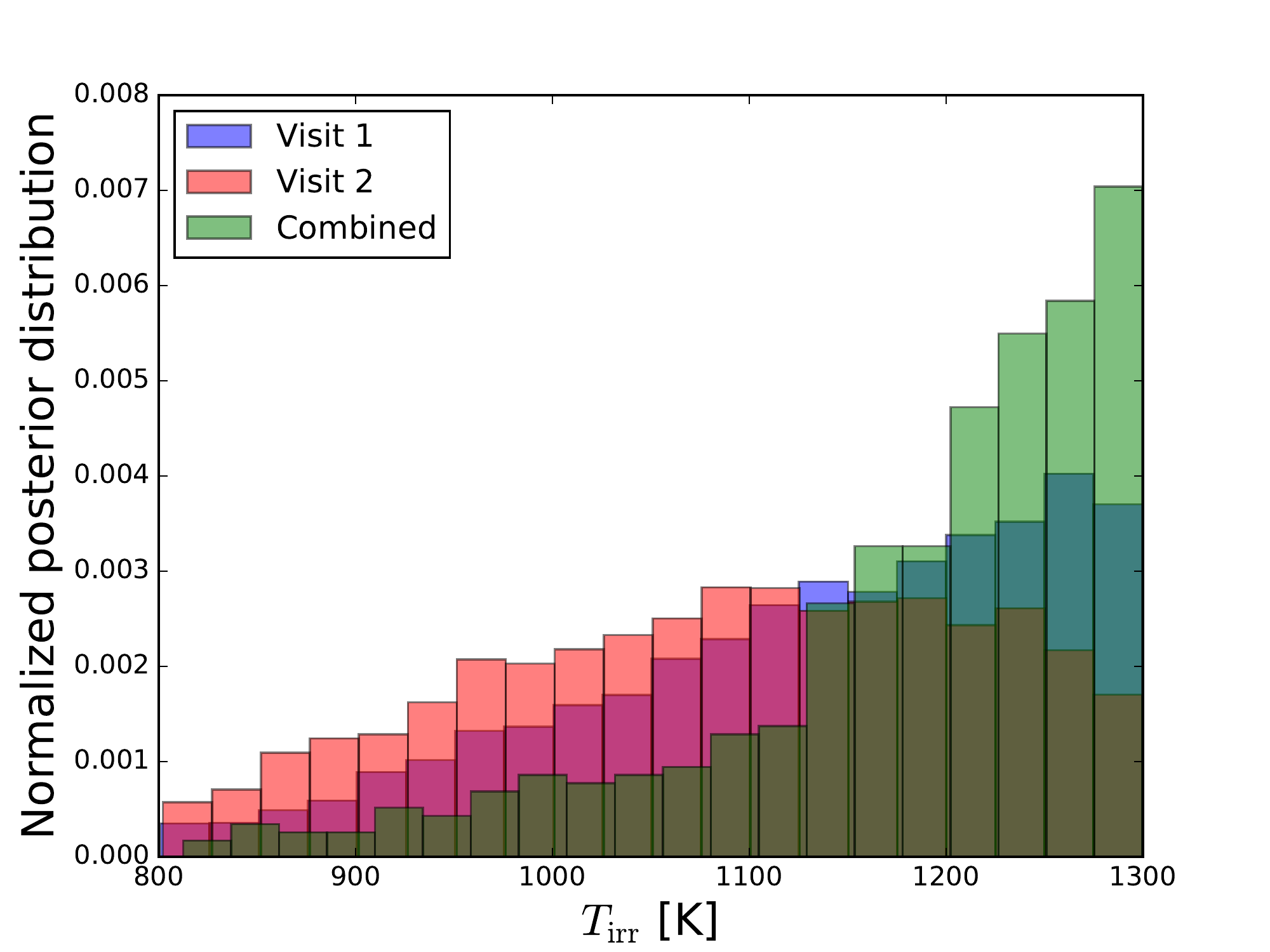}
\plotone{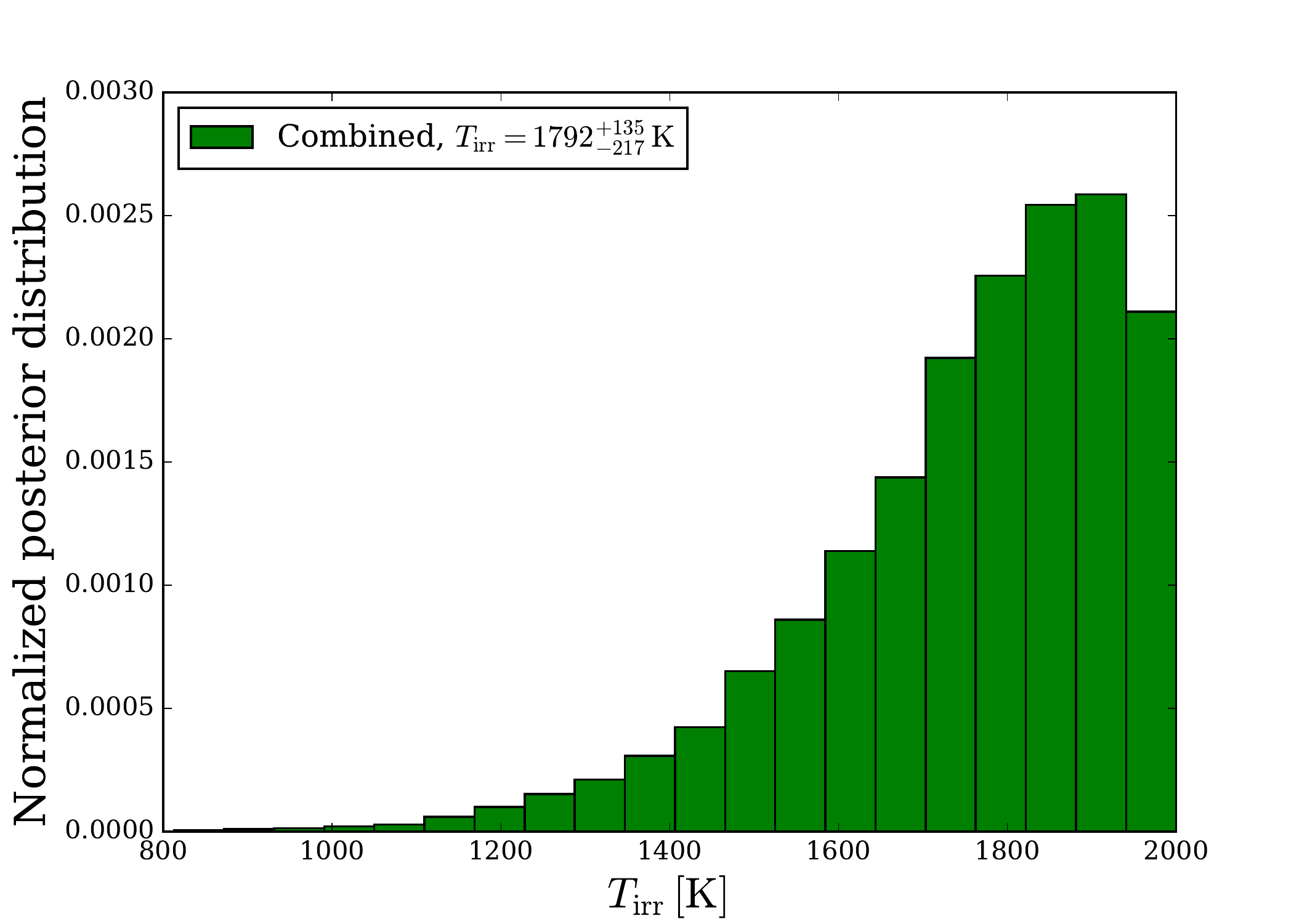}
\caption{$Top:$ Normalized posterior distributions for the water abundance in the first visit (blue), second visit (green) and combined (red) transmission spectrum of HAT-P-38 b. The vertical dashed lines separates the low-metallicity mode (mixing ratio $<10^{-2}$) from the high-metallicity one (mixing ratio $>10^{-2}$), as discussed in Section \ref{twovisits}. \textit{Center}: Marginalized posterior distribution for the $T_\mathrm{irr}$ on the combined spectrum of HAT-P-38 b, with no point rejected, and with a prior on $T_\mathrm{irr}$ going from 800 to 2000 K. The posterior for the combined spectrum is cut at the same value of the separate visits priors for clarity. \textit{Bottom}: Complete marginalized posterior distribution for $T_\mathrm{irr}$ on the combined spectrum of HAT-P-38 b, with no point rejected, together with the 15.9\%, 50\%, and 84.1\% percentiles.}
\label{waterab}
\end{figure}

Because of the difference in the visits of HAT-P-38 b, exploration retrievals are at first performed for each visit, separately. The results are presented in Figure \ref{waterab}, top panel, where the retrieved water abundances are shown for the first and second visit and for the combined transmission spectrum. The high-abundance parts of the distributions, on the right of the vertical dashed line at $\log \mathrm{H}_2\mathrm{O} = -2$, imply a $\gtrsim 100 \times$ solar metallicity, hard to explain given current giant planet composition models \citep[e.g.][]{thorngren2016}.

Lower abundances are at about $3 \sigma$ agreement for the two visits. The low-abundance part of the posterior distribution for the combined visit lies in between the two individual visits, while the high-abundance part of this posterior distribution is more peaked than the two visits taken individually. Moreover, the posterior distributions for $T_\mathrm{irr}$, shown in Figure \ref{waterab}, central panel, show that the retrievals for the first visit and for the combined spectrum move towards too high temperatures, given the \teq~$\sim 1080$ K found by \cite{hellier2012}. To find the peak of the $T_\mathrm{irr}$ posterior distribution of the combined spectrum, it is necessary to broaden the uniform distribution prior in Table \ref{priorsretrievals} from 800 to 2000 K. The average value of the posterior is $\sim 1800$ K. The complete posterior distribution for the combined spectrum, together with the related 15.9\%, 50\%, and 84.1\% percentiles, are shown for clarity in a separate panel (lowest in the Figure).

\begin{figure}
\epsscale{1.1}
\plotone{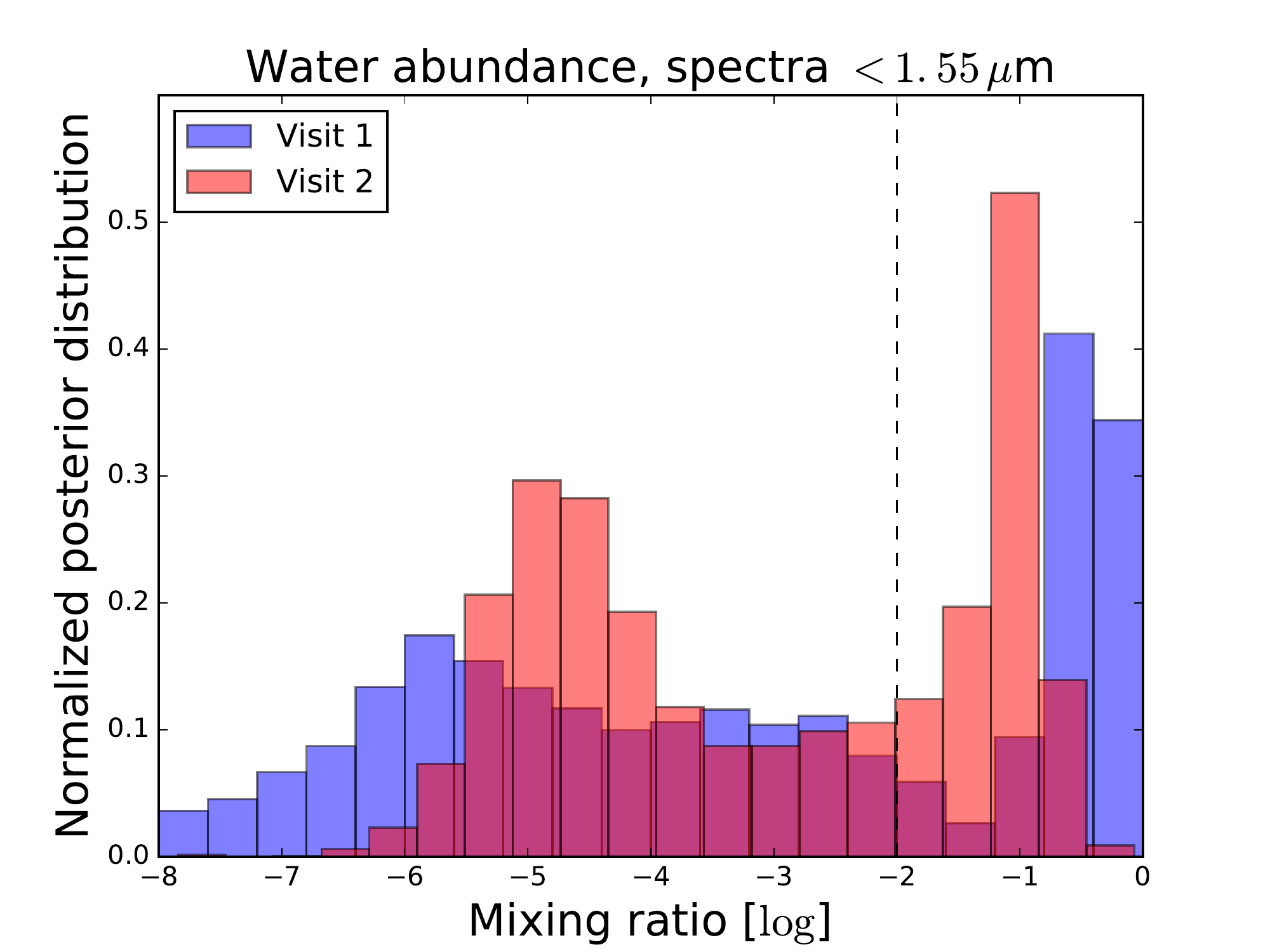}
\caption{Normalized posterior distribution for the two visits, where the part of the spectra with $\lambda > 1.55 \mu$m was removed. The black dashed lines separates the two modes found by the retrievals. Abundances larger than $\log 2$ are difficult to explain given our knowledge of Jupiter-like exoplanets.}
\label{waterab_chopped}
\end{figure}

The difference in the water abundances and the too high $T_\mathrm{irr}$ are driven by the redder channels of the two visits. To test this, we cut the two transmission spectra at wavelengths larger than 1.55 $\mu$m and repeat the separate retrievals. As shown in Figure \ref{waterab_chopped}, the water abundances are now in agreement.

Following this comparison and because we find no indication of red noise or instrumental artifacts affecting any of the two visits (section \ref{spectrotr}), we choose to analyze the combined spectrum from which the wavelengths larger than $1.55 \, \mu$m are removed. In the following sections, the complete retrievals for this latter spectrum are presented.

\begin{table}
\begin{center}
\caption{Priors for the retrievals. $\mathcal{U}(a, b)$ denotes a uniform distribution between $a$ and $b$.}
\label{priorsretrievals}
\begin{tabular}{l|l}
\hline
$T_\mathrm{irr}$ [K]                                 & $\mathcal{U}(800, 1300)$              \\
$\log (\kappa_\mathrm{IR})$ [cm$^2$g$^{-1}$]         & 0.03 (fixed)                          \\
$\log \gamma_1$                                      & 1 (fixed)                             \\
$[\mathrm{Fe/H}]$ [$\times$ solar] $^{(a)}$          & $\mathcal{U}(10^{-4},10^3)$           \\
$\log (C/O)^{(b)}$                                   & $\mathcal{U}(-2,2)$                   \\
Quench pressure $P_\mathrm{q} (\mathrm{C}$) [log bar]        & $-6$  (fixed)                    \\
Quench pressure $P_\mathrm{q} (\mathrm{N}$) [log bar]        &$-6$  (fixed)                    \\
Rayleigh Haze $\sigma_0$ [$\sigma_{\mathrm{H}_2}$]   & 0  (fixed)                             \\
Rayleigh Haze $\gamma$                               & $-4$ (fixed)                           \\
Cloud $P_c$ [log bar]                                & $\mathcal{U}(-7, 2.5)$                \\
$R_p$ scaling factor                                 & $\mathcal{U}(0.5, 1.5)$               \\
H$_2$O abundance [$\log$ mixing ratio]$^{(c)}$     & $\mathcal{U}(-12, 0)$                   \\
CH$_4$ abundance [$\log$ mixing ratio]$^{(c)}$     & $\mathcal{U}(-12, 0)$                   \\
\hline
\end{tabular}
\end{center}
\begin{tablenotes}\footnotesize
    \item $^{(a)}$ Jump parameter for the retrieval under the assumption of chemical equilibrium; fixed to solar otherwise. $^{(b)}$ Jump parameter  for the retrieval under the assumption of chemical equilibrium; fixed to solar (-0.26) otherwise. $^{(c)}$ Jump parameter for the retrieval without the assumption of chemical equilibrium; fixed to 0 (in $\log$ units) otherwise.
\end{tablenotes}
\end{table}

\begin{figure*}[tb]
\epsscale{0.55}
\plotone{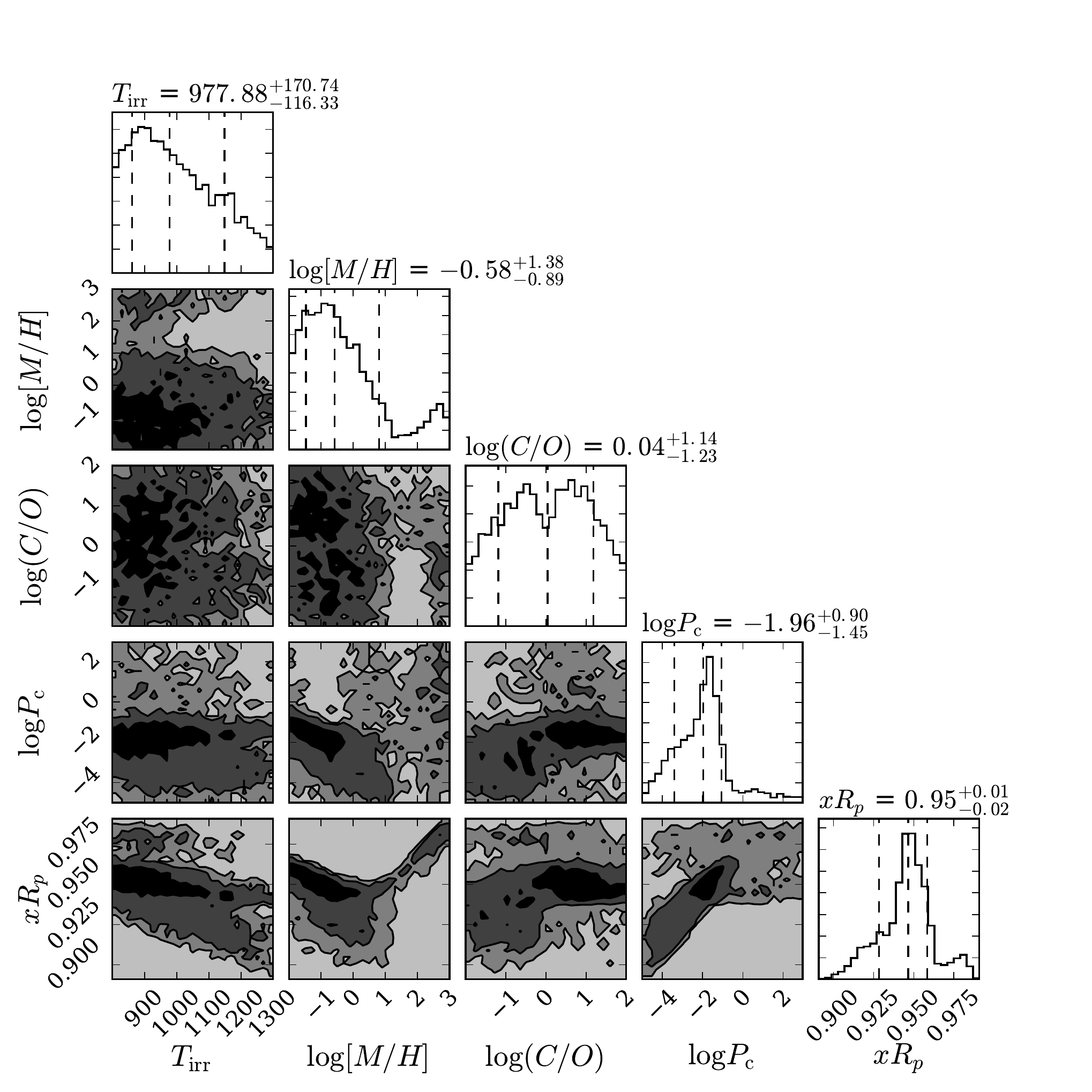}
\plotone{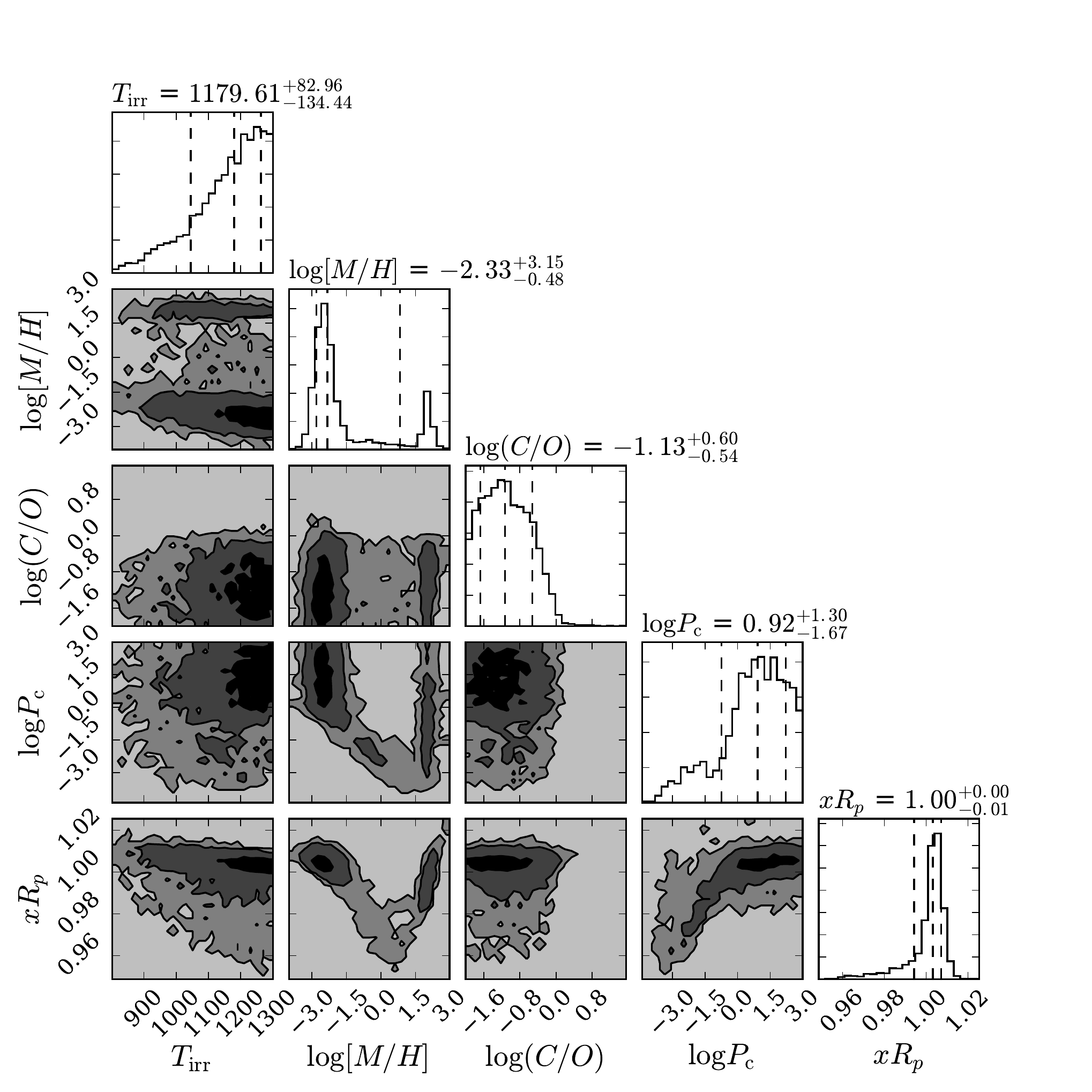}
\caption{Marginalized posterior distributions and correlation plots resulting from the retrieval on WASP-67 b (left) and HAT-P-38 b (right) assuming chemical equilibrium. Above each histogram, the 15.9\%, 50\%, and 84.1\% percentiles are reported.}
\label{pyramid_equilibrium}
\end{figure*}

\begin{figure*}[tb]
\epsscale{0.55}
\plotone{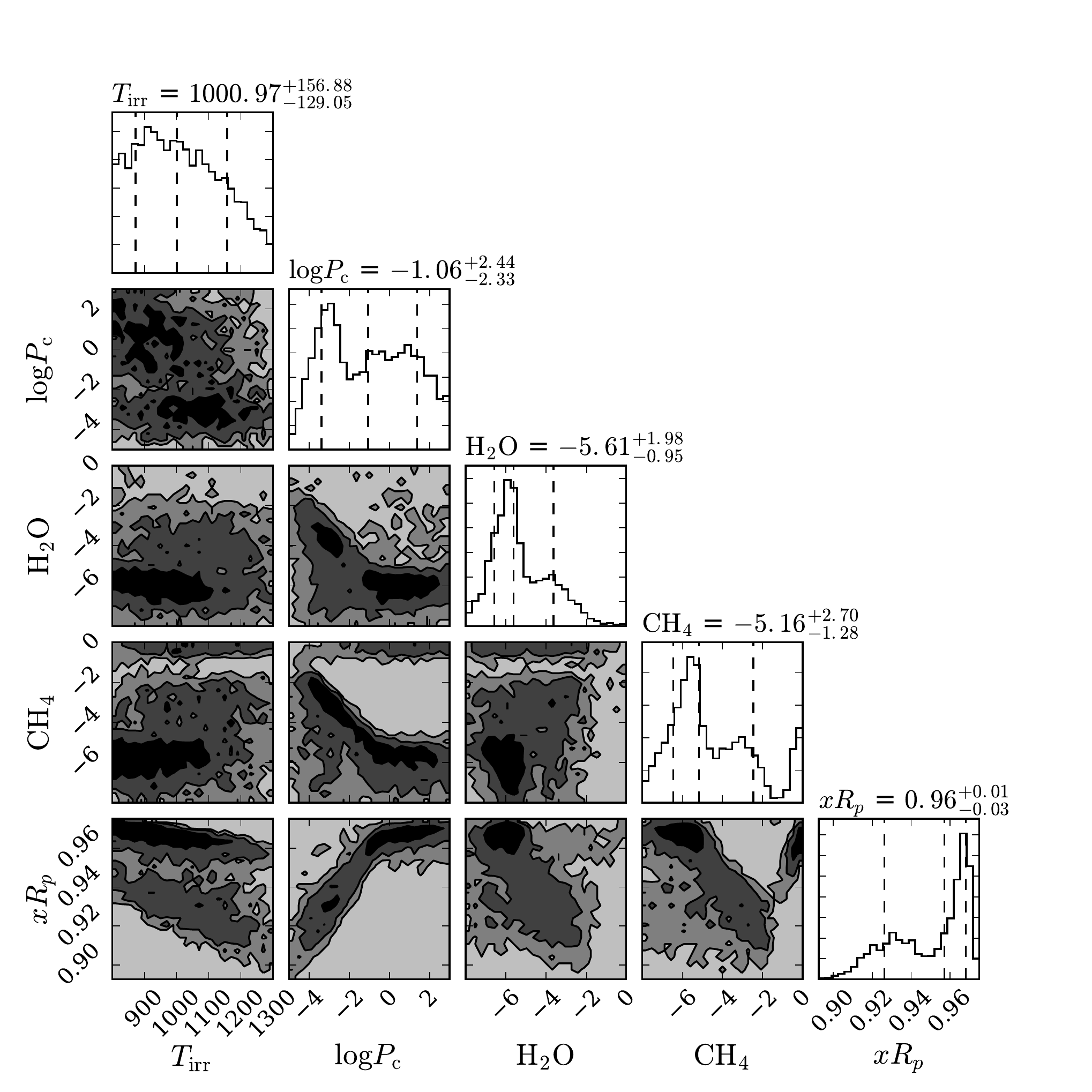}
\plotone{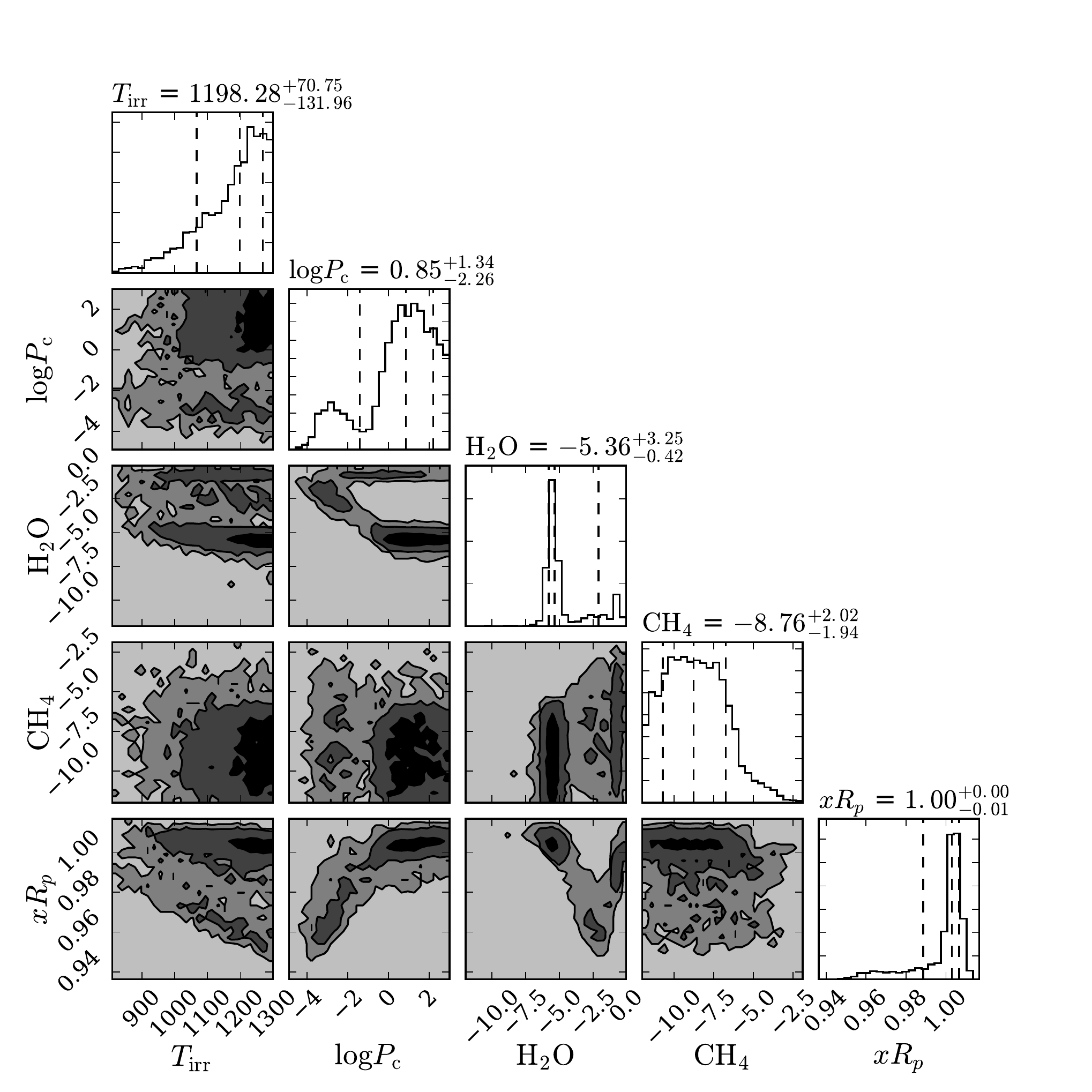}
\caption{Marginalized posterior distributions and correlation plots resulting from the retrieval on WASP-67 b (left) and HAT-P-38 b (right) without assuming chemical equilibrium. Percentiles are reported as in Figure \ref{pyramid_equilibrium}.}
\label{pyramid_noneq}
\end{figure*}

\begin{table}
\begin{center}
\caption{Reduced $\chi^2$, log-likelihood and BIC of the best-fit retrieved atmospheric models.}
\label{res_retr}
\begin{tabular}{cr|l}
\hline
    & Chemical equilibrium & Free abundances\\
\hline \hline
\underline{WASP-67 b} & & \\
$\tilde{\chi}^2$    & 1.72 & 1.62 \\
$\ln \mathcal{L}$   & -7.44 & -7.02 \\
BIC & 28.73 & 27.91 \\
\underline{HAT-P-38 b} & & \\

$\tilde{\chi}^2$    & 2.02 & 1.99 \\
$\ln \mathcal{L}$   & -8.73 & -8.74 \\
BIC & 31.00 & 31.01\\
\hline
\end{tabular}
\end{center}
\end{table}

\subsection{With chemical equilibrium}
The reduced $\chi^2$, log-likelihood and Bayesian Information Criterion (BIC) of the best-fit model from the retrieval are reported in Table \ref{res_retr}. The best likelihood solutions and the 1-- and $2\sigma$ confidence regions from the retrievals are overplotted to the spectra of each target in Figure \ref{spectraconfidence}. In the panel presenting HAT-P-38 b, the points at wavelength larger than $1.55 \, \mu$m are not considered in the retrieval, as discussed in Section \ref{twovisits}. In Figure \ref{pyramid_equilibrium}, the marginalized posterior distributions and correlation plots for both targets are shown.

The posterior distributions are driven by the muted absorption features. For the two targets, the posteriors for \tirr\ are in agreement at the $1\sigma$ level, even if slightly colder solutions are preferred for WASP-67 b. Colder solutions correspond to lower scale heights, which allow to fit for weaker absorption features. No particular physical meaning has to be attributed to these results.

Bimodal posterior distributions for [M/H] are found for both targets. The two modes separate a region with [M/H]$\gtrsim 30 \times$ (higher mean molecular weight) and $\lesssim 30 \times$ (lower mean molecular weight) solar metallicity. The low-mean molecular weight solutions are favored by our understanding of giant planets composition \citep[e.g.][]{thorngren2016}, where the retrievals recover the [M/H]-$P_\mathrm{c}$ degeneracy. For such solutions, we observe mainly $\sim 0.01-10 \times$ solar solutions for WASP-67 b and the placement of a cloud deck at higher altitude for WASP-67 b ($\gtrsim 1$ mbar at $1 \sigma$) than for HAT-P-38 b ($\gtrsim 150$ mbar at $1 \sigma$). 

As expected, WFC3 observations do not constrain the C/O ratios. High values of C/O ratios are allowed for by WASP-67 b. This is due to the point at 1.6 $\mu$m in the spectrum, which is however isolated. The transit depth at this particular wavelength is found to be dependent on the binning during the reduction (section \ref{observations}) and we therefore warn it has to be considered with caution. We finally remark the correlation between the physical retrieved parameters and the scaling of the planetary radius, reflecting the uncertainty of the reference pressure for the computation of the spectrum. In addition, the scaling of the planetary radius and the cloud top pressure are correlated, as a variation in the pressure where the optical depth goes to zero can be compensated by changing the reference radius for the planet \citep[e.g.][]{lecavalierdesetangs2008,heng2017}.

\subsection{Without chemical equilibrium}
The reduced $\chi^2$, log-likelihood and BIC of the best-fit model from this retrieval are reported in Table \ref{res_retr}. As the best-likelihood models are indistinguishable by eye from the previous retrievals, we refer to Figure \ref{spectraconfidence} for the best-likelihood models and confidence regions. Relaxing the requirement of chemical equilibrium by fixing the metallicity and C/O ratio to solar values ($[\mathrm{M/H}] = 0, \, \mathrm{C/O} = -0.26$) enables retrieving the abundances of water and methane. Plot \ref{pyramid_noneq} shows the marginalized posterior distributions and correlation plots for the retrieved parameters.

The water abundance for both targets shows a bimodal distribution, which confirms the bimodal posteriors retrieved for the metallicity. Strong correlations are found between the cloud top pressure and the water abundance in the low-abundance mode, as for the metallicity. Solar and sub-solar water abundances are retrieved for both targets, with lower abundances allowed for WASP-67 b than for HAT-P-38 b (mixing ratio $\gtrsim 3 \times 10^{-7}$ against $\gtrsim 2 \times 10^{-6}$ at $1\sigma$; the solar water abundance corresponds to $\sim 10^{-4}$, \citealp{lodders2009}). High water abundance solutions, corresponding to the high-metallicity solutions of the retrieval with chemical equilibrium, are retrieved for HAT-P-38 b. Such solutions correspond to implausibly water-rich atmospheres, which would be hardly compatible with current models of giant planet composition. However, as these solutions are allowed for by the observations, they are conservatively reported.

Large uncertainties for the methane abundance are retrieved in both cases. As expected, given the point at about $1.6 \, \mu$m already discussed for WASP-67 b, the uncertainty for methane is larger for this target, Over all, our abundances are in agreement with those found by \cite{tsiaras2017}, who performed retrievals on a large number of planets and allowed mixing rations down to $10^{-10}$. As \citeauthor{tsiaras2017} present only one spectrum per HAT-P-38 b, however, we are not able to conclude whether they also find the same difference between the two visits.

\section{Discussion}\label{future}
The measured spectra of WASP-67 b and HAT-P-38 b indicate that these twin planets, with nearly identical temperatures, gravities, and host stars, have very different atmospheric properties. Indeed, atmospheric metallicity is another unavoidable parameter to consider, in order to determine the composition of clouds which can affect a transmission spectrum \citep{viisscher2006,visscher2010,morley2015}. First-order expectations on the condensed species are usually obtained from atmospheric models with solar composition. In this scenario, the P-T structure of the atmosphere is compared to Clausius-Clapeyron curves for the dominant species, determining e.g. the base pressure and extent of clouds \citep{sanchezlavega2004}. 

\begin{figure}[tb]
\epsscale{1.3}
\plotone{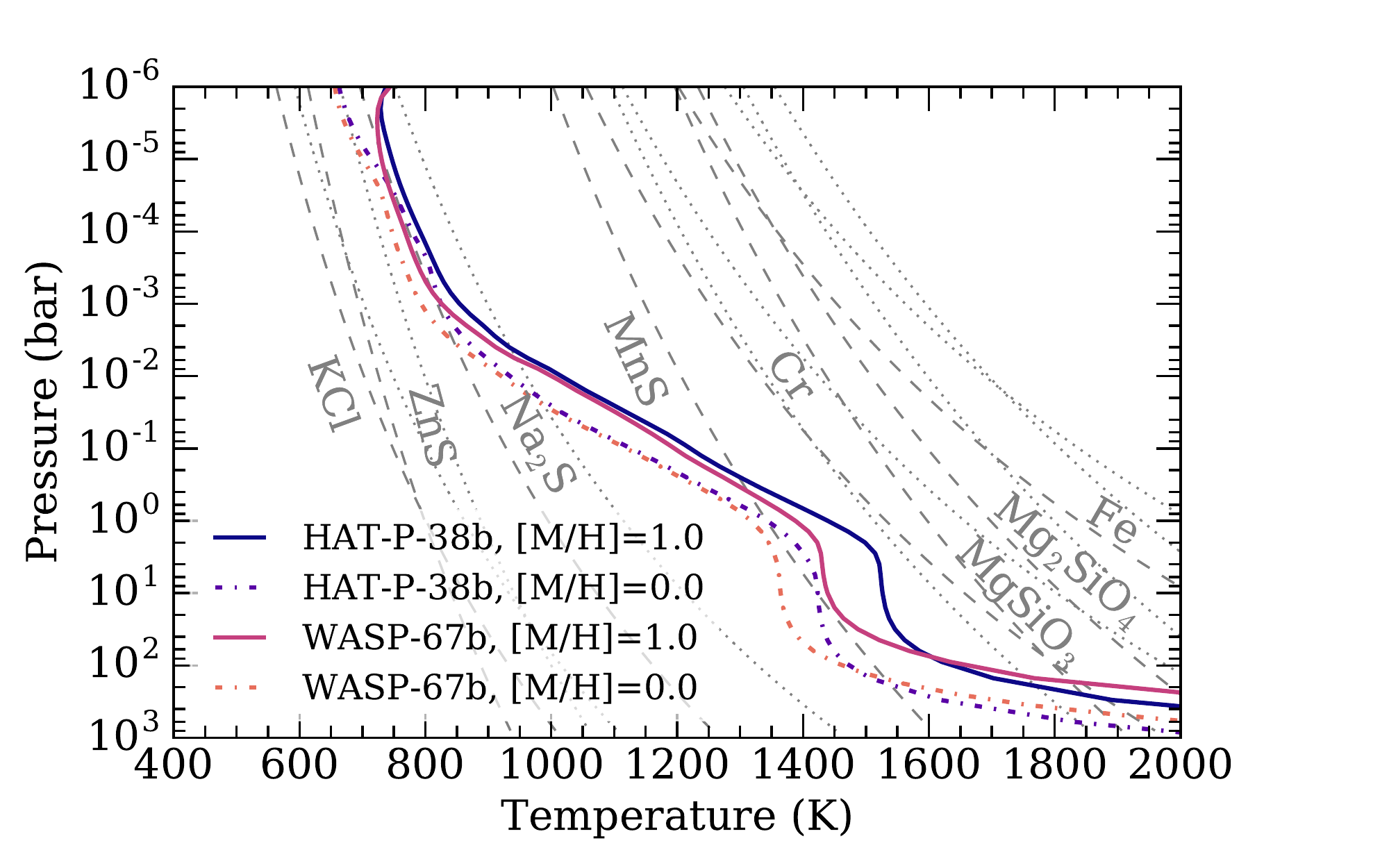}
\caption{Difference in the P-T profiles of WASP-67 b and HAT-P-38 b from $1\times$ (dashed lines) to $10\times$ (continuous lines) solar metallicity. Given the mass-metallicity relation, the P-T profile of WASP-67 b should be closer to the dashed line and the one of HAT-P-38 b closer to the continuous line.}
\label{met10x}
\end{figure}

Focusing on two planets with nearly equal temperature and gravity, we can isolate the role of metallicity in shaping transmission spectra. From current constraints of the planet mass-metallicity relation \citep[e.g.][and references therein]{wakeford2017}, we expect both HAT-P-38 b and WASP-67 b to lie in a metallicity range of $1\times$ to $10\times$ solar. Given the smaller mass of HAT-P-38 b with respect to WASP-67 b (0.27 against 0.42 M$_\mathrm{J}$), we also expect the former to be slightly more metal-rich than the latter. A higher metallicity for HAT-P-38 b than for WASP-67 b would cause a shift of both the P-T profiles of Figure \ref{ptprofs} and of the condensation curves on the temperature axis, as shown in Figure \ref{met10x}. As a result, alkali-bearing condensates such as MnS and Na$_2$S would form deeper in the atmosphere for WASP-67 b than for HAT-P-38 b, obscuring to a larger extent the pressure regime probed by transmission spectroscopy. The comparison of these two planets points, therefore, to the possibility of constraining the composition of exoplanet atmospheres by using the muting of spectral features due to aerosols.

Current observations in the near-infrared, however, cannot effectively constrain $[\mathrm{M/H}]$ nor cloud species. Retrievals rely on a generic gray cloud top pressure and the role of metallicity in shaping transmission spectra cannot be clearly elucidated. Advances in this direction need a broader wavelength coverage. Optical wavelengths are already accessible by present instrumentation and allow constraining the kind of atmospheric aerosols, especially hazes. {\em HST}/STIS would be particularly suitable for distinguishing different scattering particle sizes and slopes \citep[e.g.][]{lecavalierdesetangs2008}, in order to constrain the scattering parameters $\sigma_0$ and $\gamma$ which do not affect G141 observations. By probing alkali lines, observations in the visible would also constrain the temperature structure of the atmospheres above 1 mbar \citep[e.g.][]{vidal-madjar2011,vidal-madjar2011corr}, complementing near-IR observations which can only probe regions between $\sim 1-100$ mbar. A better constraint on the P-T profiles would allow more detailed retrievals on the atmospheric parameters, relaxing the assumption of nearly isothermal atmosphere, and a better insight on the base pressure of the cloud species.

The mid-IR ($5-28 \mu$m) is another spectral region of crucial interest. Observations in this band would allow the distinction of condensates vibrational modes and therefore of different condensate species \citep{wakeford2015}. JWST/MIRI will be able to shed light on this aspect.\\

Giant planets are expected to retain much of their primordial composition after their formation and evolution. The abundances we observe are therefore the result of the composition of the protoplanetary disk where they formed, of their initial location in the disk, and of the accreted materials during the evolution history \citep[e.g.][and references therein]{pollack1996,boss1997,alibert2005,madhusudhan2011,helling2014,kreidberg2014,mordasini2016,thorngren2016}. The different characteristics of the two atmospheres keep trace of their different formation and evolution environments. In this context, the measurement of the scattering signature in the visible with STIS would enable breaking the cloud-metallicity degeneracy in the retrievals \citep[e.g.][]{benneke2012}, improving our knowledge of the mass--metallicity relation for giant planets. The resulting constraints on their formation and evolutionary history are particularly valuable for HAT-P-38 b, as little sampling is currently available for Saturn-mass bodies.

Population synthesis models which follow the entire planets' history and return their composition at the end of their migration were explored in few studies, and strong observational constraints on the outcome of the models are expected with instruments such as JWST and the ESA candidate mission ARIEL \citep{pace2016}. This is one more reason to pursue follow-up observations in both visible and mid-IR of WASP-67 b and HAT-P-38 b, in order to access vital information for planet evolutionary models.

We remark in this place the importance of compared analysis. Similar reductions of the spectra and light curve fits were performed independently by different members of our team. The comparison of the results ensured the robustness of the analysis. This was useful both in assessing the importance of the point at 1.6 $\mu$m for WASP-67 b and in comparing the two visits of HAT-P-38 b.

\section{Conclusions}\label{conclusions}
We present WFC3/G141 transmission spectroscopy observations of the short-period giant planets WASP-67 b and HAT-P-38 b. While having nearly equal irradiation level (\teq\ $\sim 1050 \, \mathrm{K}$) and gravity ($g \sim 10 \, \mathrm{m \, s}^{-1})$, thought to be the main parameters determining cloud formation and extent other than composition, the spectra of these two planets are remarkably different.

A slope is observed in the spectrum of the first visit of HAT-P-38 b. Despite the two visits agreeing at $1 \sigma$ for most channels, the difference in the wavelengths larger than $1.55 \, \mu$m is large enough to cause unrealistic results in the retrievals on their combined spectrum. Several tests are performed to exclude the presence of red noise in the first visit as in the second. We cannot identify the source of the slope, which we attribute to an instrumental artifact, and we exclude the last channels from our analysis.

The abundances we retrieve for both planets are in agreement with those found by \cite{tsiaras2017}, who perform retrievals large sample of exoplanet spectra. We are unable to conclude whether these authors find a difference in the two visits of HAT-P-38 b, as they present only a spectrum for this planet. We detect water in both planets and attribute the different significance of their water peaks at 1.4 $\mu$m both to a muting effect due to obscuring clouds and to different formation and accretion histories. Our analysis also recovers a correlation between the retrieved abundance of water and the cloud base altitude. 

Muted water absorption can be explained by sub-solar water abundance, solar water abundance with clouds, or very high water abundance. Given the masses of these planets, however, both very low and very high metallicity solutions are disfavored. For the solar and sub-solar abundance, we recover the cloud top-pressure--metallicity degeneracy. We suggest that the different atmospheric metallicity of the planets, likely separated by about an order of magnitude, affects the base pressure of alkali clouds, as expected from aerosol models in literature. As G141, near-IR observations alone are not able to constrain metallicity nor cloud top pressure, optical ({\em HST}/STIS) and mid-IR (JWST and ARIEL) observations are discussed in their potential to solve the degeneracy.

Follow-up observations are particularly important for WASP-67 b and HAT-P-38 b, as their comparative study could unveil vital information for the understanding of aerosols in giant planet atmospheres.



\acknowledgments

\small{Based on observations made with the NASA/ESA Hubble Space Telescope, obtained from the data archive at the Space Telescope Science Institute. STScI is operated by the Association of Universities for Research in Astronomy, Inc. under NASA contract NAS 5-26555. These observations are associated with program GO 14260.}



Facilities: \facility{{\em HST}(WFC3)}.



\bibliography{biblio}

\begin{thebibliography}{}
\expandafter\ifx\csname natexlab\endcsname\relax\def\natexlab#1{#1}\fi

\bibitem[{{Alibert} {et~al.}(2005){Alibert}, {Mordasini}, {Benz}, \&
  {Winisdoerffer}}]{alibert2005}
{Alibert}, Y., {Mordasini}, C., {Benz}, W., \& {Winisdoerffer}, C. 2005, \aap,
  434, 343

\bibitem[{{Baker} \& {Matthews}(2001)}]{baker2001}
{Baker}, S., \& {Matthews}, I. 2001, Proceedings of the 2001 IEEE Conference on
  Computer Vision and Pattern Recognition, Volume 1, Pages 1090--1097

\bibitem[{{Benneke} \& {Seager}(2012)}]{benneke2012}
{Benneke}, B., \& {Seager}, S. 2012, \apj, 753, 100

\bibitem[{{Berta} {et~al.}(2012){Berta}, {Charbonneau}, {D{\'e}sert},
  {Miller-Ricci Kempton}, {McCullough}, {Burke}, {Fortney}, {Irwin}, {Nutzman},
  \& {Homeier}}]{berta2012}
{Berta}, Z.~K., {Charbonneau}, D., {D{\'e}sert}, J.-M., {et~al.} 2012, \apj,
  747, 35

\bibitem[{{Boss}(1997)}]{boss1997}
{Boss}, A.~P. 1997, Science, 276, 1836

\bibitem[{Brooks \& Gelman(1998)}]{brooks1998}
Brooks, S.~P., \& Gelman, A. 1998, Journal of Computational and Graphical
  Statistics, 7, 434

\bibitem[{{Buchner} {et~al.}(2014){Buchner}, {Georgakakis}, {Nandra}, {Hsu},
  {Rangel}, {Brightman}, {Merloni}, {Salvato}, {Donley}, \&
  {Kocevski}}]{buchner2014}
{Buchner}, J., {Georgakakis}, A., {Nandra}, K., {et~al.} 2014, \aap, 564, A125

\bibitem[{{Burrows} {et~al.}(2000){Burrows}, {Marley}, \&
  {Sharp}}]{burrows2000}
{Burrows}, A., {Marley}, M.~S., \& {Sharp}, C.~M. 2000, \apj, 531, 438

\bibitem[{{Charbonneau} {et~al.}(2002){Charbonneau}, {Brown}, {Noyes}, \&
  {Gilliland}}]{charbonneau2002}
{Charbonneau}, D., {Brown}, T.~M., {Noyes}, R.~W., \& {Gilliland}, R.~L. 2002,
  \apj, 568, 377

\bibitem[{{Claret}(2000)}]{claret2000}
{Claret}, A. 2000, \aap, 363, 1081

\bibitem[{{Cubillos} {et~al.}(2016){Cubillos}, {Harrington}, {Lust}, {Foster},
  {Stemm}, {Loredo}, {Stevenson}, {Campo}, {Hardin}, \& {Hardy}}]{cubillos2016}
{Cubillos}, P., {Harrington}, J., {Lust}, N., {et~al.} 2016, {MC3: Multi-core
  Markov-chain Monte Carlo code}, Astrophysics Source Code Library,
  ascl:1610.013

\bibitem[{{Deming} {et~al.}(2013){Deming}, {Wilkins}, {McCullough}, {Burrows},
  {Fortney}, {Agol}, {Dobbs-Dixon}, {Madhusudhan}, {Crouzet}, {Desert},
  {Gilliland}, {Haynes}, {Knutson}, {Line}, {Magic}, {Mandell}, {Ranjan},
  {Charbonneau}, {Clampin}, {Seager}, \& {Showman}}]{deming2013}
{Deming}, D., {Wilkins}, A., {McCullough}, P., {et~al.} 2013, \apj, 774, 95

\bibitem[{{Doyle} {et~al.}(2011){Doyle}, {Carter}, {Fabrycky}, {Slawson},
  {Howell}, {Winn}, {Orosz}, {Pr{\^s}a}, {Welsh}, {Quinn}, {Latham}, {Torres},
  {Buchhave}, {Marcy}, {Fortney}, {Shporer}, {Ford}, {Lissauer}, {Ragozzine},
  {Rucker}, {Batalha}, {Jenkins}, {Borucki}, {Koch}, {Middour}, {Hall},
  {McCauliff}, {Fanelli}, {Quintana}, {Holman}, {Caldwell}, {Still},
  {Stefanik}, {Brown}, {Esquerdo}, {Tang}, {Furesz}, {Geary}, {Berlind},
  {Calkins}, {Short}, {Steffen}, {Sasselov}, {Dunham}, {Cochran}, {Boss},
  {Haas}, {Buzasi}, \& {Fischer}}]{doyle2011}
{Doyle}, L.~R., {Carter}, J.~A., {Fabrycky}, D.~C., {et~al.} 2011, Science,
  333, 1602

\bibitem[{{Enoch} {et~al.}(2011){Enoch}, {Cameron}, {Anderson}, {Lister},
  {Hellier}, {Maxted}, {Queloz}, {Smalley}, {Triaud}, {West}, {Brown},
  {Gillon}, {Hebb}, {Lendl}, {Parley}, {Pepe}, {Pollacco}, {Segransan},
  {Simpson}, {Street}, \& {Udry}}]{enoch2011}
{Enoch}, B., {Cameron}, A.~C., {Anderson}, D.~R., {et~al.} 2011, \mnras, 410,
  1631

\bibitem[{{Feroz} {et~al.}(2009){Feroz}, {Hobson}, \& {Bridges}}]{feroz2009}
{Feroz}, F., {Hobson}, M.~P., \& {Bridges}, M. 2009, \mnras, 398, 1601

\bibitem[{{Fortney}(2005)}]{fortney2005}
{Fortney}, J.~J. 2005, \mnras, 364, 649

\bibitem[{{Fortney} {et~al.}(2008){Fortney}, {Lodders}, {Marley}, \&
  {Freedman}}]{fortney2008}
{Fortney}, J.~J., {Lodders}, K., {Marley}, M.~S., \& {Freedman}, R.~S. 2008,
  \apj, 678, 1419

\bibitem[{{Fortney} {et~al.}(2007){Fortney}, {Marley}, \&
  {Barnes}}]{fortney2007}
{Fortney}, J.~J., {Marley}, M.~S., \& {Barnes}, J.~W. 2007, \apj, 659, 1661

\bibitem[{{Fortney} {et~al.}(2003){Fortney}, {Sudarsky}, {Hubeny}, {Cooper},
  {Hubbard}, {Burrows}, \& {Lunine}}]{fortney2003}
{Fortney}, J.~J., {Sudarsky}, D., {Hubeny}, I., {et~al.} 2003, \apj, 589, 615

\bibitem[{{Fraine} {et~al.}(2014){Fraine}, {Deming}, {Benneke}, {Knutson},
  {Jord{\'a}n}, {Espinoza}, {Madhusudhan}, {Wilkins}, \&
  {Todorov}}]{fraine2014}
{Fraine}, J., {Deming}, D., {Benneke}, B., {et~al.} 2014, \nat, 513, 526

\bibitem[{Gelman \& Rubin(1992)}]{gelman1992}
Gelman, A., \& Rubin, D.~B. 1992, Statist. Sci., 7, 457

\bibitem[{{Golub} \& {van Loan}(1996)}]{golub1996}
{Golub}, G.~H., \& {van Loan}, C.~F. 1996, {Matrix computations}

\bibitem[{{Gregory}(2005)}]{gregory2005}
{Gregory}, P.~C. 2005, {Bayesian Logical Data Analysis for the Physical
  Sciences: A Comparative Approach with `Mathematica' Support} (Cambridge
  University Press)

\bibitem[{{Hartman} {et~al.}(2011){Hartman}, {Bakos}, {Sato}, {Torres},
  {Noyes}, {Latham}, {Kov{\'a}cs}, {Fischer}, {Howard}, {Johnson}, {Marcy},
  {Buchhave}, {F{\"u}resz}, {Perumpilly}, {B{\'e}ky}, {Stefanik}, {Sasselov},
  {Esquerdo}, {Everett}, {Csubry}, {L{\'a}z{\'a}r}, {Papp}, \&
  {S{\'a}ri}}]{hartman2011}
{Hartman}, J.~D., {Bakos}, G.~{\'A}., {Sato}, B., {et~al.} 2011, \apj, 726, 52

\bibitem[{{Hellier} {et~al.}(2012){Hellier}, {Anderson}, {Collier Cameron},
  {Doyle}, {Fumel}, {Gillon}, {Jehin}, {Lendl}, {Maxted}, {Pepe}, {Pollacco},
  {Queloz}, {S{\'e}gransan}, {Smalley}, {Smith}, {Southworth}, {Triaud},
  {Udry}, \& {West}}]{hellier2012}
{Hellier}, C., {Anderson}, D.~R., {Collier Cameron}, A., {et~al.} 2012, \mnras,
  426, 739

\bibitem[{{Helling} {et~al.}(2014){Helling}, {Woitke}, {Rimmer}, {Kamp}, {Thi},
  \& {Meijerink}}]{helling2014}
{Helling}, C., {Woitke}, P., {Rimmer}, P.~B., {et~al.} 2014, Life, 4,
  arXiv:1403.4420

\bibitem[{{Heng} \& {Kitzmann}(2017)}]{heng2017}
{Heng}, K., \& {Kitzmann}, D. 2017, ArXiv e-prints, arXiv:1702.02051

\bibitem[{{Herdin} {et~al.}(2005){Herdin}, {Czink}, {{\``O}zcelik}, \&
  {Bonek}}]{herdin2005}
{Herdin}, M., {Czink}, N., {{\``O}zcelik}, H., \& {Bonek}, E. 2005

\bibitem[{{Horne}(1986)}]{horne1986}
{Horne}, K. 1986, \pasp, 98, 609

\bibitem[{{Husser} {et~al.}(2013){Husser}, {Wende-von Berg}, {Dreizler},
  {Homeier}, {Reiners}, {Barman}, \& {Hauschildt}}]{husser2013}
{Husser}, T.-O., {Wende-von Berg}, S., {Dreizler}, S., {et~al.} 2013, \aap,
  553, A6

\bibitem[{{Kilpatrick} {et~al.}(2017){Kilpatrick}, {Cubillos}, {Stevenson},
  {Lewis}, {Wakeford}, {Macdonald}, {Madhusudhan}, {Blecic}, {Bruno},
  {Burrows}, {Deming}, {Heng}, {Line}, {Morley}, {Parmentier}, {Tucker},
  {Valenti}, {Waldmann}, {Bean}, {Beichman}, {Fraine}, {Krick}, {Lothringer},
  \& {Mandell}}]{kilpatrick2017}
{Kilpatrick}, B.~M., {Cubillos}, P.~E., {Stevenson}, K.~B., {et~al.} 2017,
  ArXiv e-prints, arXiv:1704.07421

\bibitem[{{Knutson} {et~al.}(2014){Knutson}, {Dragomir}, {Kreidberg},
  {Kempton}, {McCullough}, {Fortney}, {Bean}, {Gillon}, {Homeier}, \&
  {Howard}}]{knutson2014}
{Knutson}, H.~A., {Dragomir}, D., {Kreidberg}, L., {et~al.} 2014, \apj, 794,
  155

\bibitem[{{Kreidberg} {et~al.}(2014{\natexlab{a}}){Kreidberg}, {Bean},
  {D{\'e}sert}, {Line}, {Fortney}, {Madhusudhan}, {Stevenson}, {Showman},
  {Charbonneau}, {McCullough}, {Seager}, {Burrows}, {Henry}, {Williamson},
  {Kataria}, \& {Homeier}}]{kreidberg2014}
{Kreidberg}, L., {Bean}, J.~L., {D{\'e}sert}, J.-M., {et~al.}
  2014{\natexlab{a}}, \apjl, 793, L27

\bibitem[{{Kreidberg} {et~al.}(2014{\natexlab{b}}){Kreidberg}, {Bean},
  {D{\'e}sert}, {Benneke}, {Deming}, {Stevenson}, {Seager}, {Berta-Thompson},
  {Seifahrt}, \& {Homeier}}]{kreidberg2014_nat}
---. 2014{\natexlab{b}}, \nat, 505, 69

\bibitem[{{Kreidberg} {et~al.}(2015){Kreidberg}, {Line}, {Bean}, {Stevenson},
  {D{\'e}sert}, {Madhusudhan}, {Fortney}, {Barstow}, {Henry}, {Williamson}, \&
  {Showman}}]{kreidberg2015}
{Kreidberg}, L., {Line}, M.~R., {Bean}, J.~L., {et~al.} 2015, \apj, 814, 66

\bibitem[{{Lecavelier Des Etangs} {et~al.}(2008){Lecavelier Des Etangs},
  {Pont}, {Vidal-Madjar}, \& {Sing}}]{lecavalierdesetangs2008}
{Lecavelier Des Etangs}, A., {Pont}, F., {Vidal-Madjar}, A., \& {Sing}, D.
  2008, \aap, 481, L83

\bibitem[{{Line} {et~al.}(2013){Line}, {Wolf}, {Zhang}, {Knutson}, {Kammer},
  {Ellison}, {Deroo}, {Crisp}, \& {Yung}}]{line2013}
{Line}, M.~R., {Wolf}, A.~S., {Zhang}, X., {et~al.} 2013, \apj, 775, 137

\bibitem[{{Lodders}(1999)}]{lodders1999}
{Lodders}, K. 1999, \apj, 519, 793

\bibitem[{{Lodders} {et~al.}(2009){Lodders}, {Palme}, \& {Gail}}]{lodders2009}
{Lodders}, K., {Palme}, H., \& {Gail}, H.-P. 2009, Landolt B{\"o}rnstein,
  arXiv:0901.1149

\bibitem[{{Madhusudhan} {et~al.}(2014{\natexlab{a}}){Madhusudhan}, {Amin}, \&
  {Kennedy}}]{madhusudhan2014_2}
{Madhusudhan}, N., {Amin}, M.~A., \& {Kennedy}, G.~M. 2014{\natexlab{a}},
  \apjl, 794, L12

\bibitem[{{Madhusudhan} {et~al.}(2014{\natexlab{b}}){Madhusudhan}, {Crouzet},
  {McCullough}, {Deming}, \& {Hedges}}]{madhusudhan2014}
{Madhusudhan}, N., {Crouzet}, N., {McCullough}, P.~R., {Deming}, D., \&
  {Hedges}, C. 2014{\natexlab{b}}, \apjl, 791, L9

\bibitem[{{Madhusudhan} {et~al.}(2011){Madhusudhan}, {Mousis}, {Johnson}, \&
  {Lunine}}]{madhusudhan2011}
{Madhusudhan}, N., {Mousis}, O., {Johnson}, T.~V., \& {Lunine}, J.~I. 2011,
  \apj, 743, 191

\bibitem[{{Mancini} {et~al.}(2014){Mancini}, {Southworth}, {Ciceri}, {Calchi
  Novati}, {Dominik}, {Henning}, {J{\o}rgensen}, {Korhonen}, {Nikolov},
  {Alsubai}, {Bozza}, {Bramich}, {D'Ago}, {Figuera Jaimes}, {Galianni}, {Gu},
  {Harps{\o}e}, {Hinse}, {Hundertmark}, {Juncher}, {Kains}, {Popovas}, {Rabus},
  {Rahvar}, {Skottfelt}, {Snodgrass}, {Street}, {Surdej}, {Tsapras}, {Vilela},
  {Wang}, \& {Wertz}}]{mancini2014}
{Mancini}, L., {Southworth}, J., {Ciceri}, S., {et~al.} 2014, \aap, 568, A127

\bibitem[{{Mandel} \& {Agol}(2002)}]{mandelagol2002}
{Mandel}, K., \& {Agol}, E. 2002, \apjl, 580, L171

\bibitem[{{Marley} {et~al.}(1996){Marley}, {Saumon}, {Guillot}, {Freedman},
  {Hubbard}, {Burrows}, \& {Lunine}}]{marley1996}
{Marley}, M.~S., {Saumon}, D., {Guillot}, T., {et~al.} 1996, Science, 272, 1919

\bibitem[{{McCullough} \& {MacKenty}(2012)}]{mccullough2012}
{McCullough}, P., \& {MacKenty}, J. 2012, {Considerations for using Spatial
  Scans with WFC3}, Tech. rep.

\bibitem[{{McKay} {et~al.}(1989){McKay}, {Pollack}, \& {Courtin}}]{mckay1989}
{McKay}, C.~P., {Pollack}, J.~B., \& {Courtin}, R. 1989, \icarus, 80, 23

\bibitem[{{Mordasini} {et~al.}(2016){Mordasini}, {van Boekel}, {Molli{\`e}re},
  {Henning}, \& {Benneke}}]{mordasini2016}
{Mordasini}, C., {van Boekel}, R., {Molli{\`e}re}, P., {Henning}, T., \&
  {Benneke}, B. 2016, \apj, 832, 41

\bibitem[{{Morley} {et~al.}(2012){Morley}, {Fortney}, {Marley}, {Visscher},
  {Saumon}, \& {Leggett}}]{morley2012}
{Morley}, C.~V., {Fortney}, J.~J., {Marley}, M.~S., {et~al.} 2012, \apj, 756,
  172

\bibitem[{{Morley} {et~al.}(2015){Morley}, {Fortney}, {Marley}, {Zahnle},
  {Line}, {Kempton}, {Lewis}, \& {Cahoy}}]{morley2015}
---. 2015, \apj, 815, 110

\bibitem[{{Pace} {et~al.}(2016){Pace}, {Micela}, \& {Ariel Team}}]{pace2016}
{Pace}, E., {Micela}, G., \& {Ariel Team}. 2016, \memsai, 87, 214

\bibitem[{{Parviainen}(2015)}]{parviainen2015}
{Parviainen}, H. 2015, \mnras, 450, 3233

\bibitem[{{Pirzkal} {et~al.}(2016){Pirzkal}, {Ryan}, \&
  {Brammer}}]{pirzkal2016}
{Pirzkal}, N., {Ryan}, R., \& {Brammer}, G. 2016, {Trace and Wavelength
  Calibrations of the WFC3 G102 and G141 IR Grisms}, Tech. rep.

\bibitem[{{Pollack} {et~al.}(1996){Pollack}, {Hubickyj}, {Bodenheimer},
  {Lissauer}, {Podolak}, \& {Greenzweig}}]{pollack1996}
{Pollack}, J.~B., {Hubickyj}, O., {Bodenheimer}, P., {et~al.} 1996, \icarus,
  124, 62

\bibitem[{{S{\'a}nchez-Lavega} {et~al.}(2004){S{\'a}nchez-Lavega},
  {P{\'e}rez-Hoyos}, \& {Hueso}}]{sanchezlavega2004}
{S{\'a}nchez-Lavega}, A., {P{\'e}rez-Hoyos}, S., \& {Hueso}, R. 2004, American
  Journal of Physics, 72, 767

\bibitem[{{Sato} {et~al.}(2012){Sato}, {Hartman}, {Bakos}, {B{\'e}ky},
  {Torres}, {Latham}, {Kov{\'a}cs}, {Csubry}, {Penev}, {Noyes}, {Buchhave},
  {Quinn}, {Everett}, {Esquerdo}, {Fischer}, {Howard}, {Johnson}, {Marcy},
  {Sasselov}, {Szklen{\'a}r}, {L{\'a}z{\'a}r}, {Papp}, \&
  {S{\'a}ri}}]{sato2012}
{Sato}, B., {Hartman}, J.~D., {Bakos}, G.~{\'A}., {et~al.} 2012, \pasj, 64, 97

\bibitem[{{Seager} {et~al.}(2005){Seager}, {Richardson}, {Hansen}, {Menou},
  {Cho}, \& {Deming}}]{seager2005}
{Seager}, S., {Richardson}, L.~J., {Hansen}, B.~M.~S., {et~al.} 2005, \apj,
  632, 1122

\bibitem[{{Seager} \& {Sasselov}(2000)}]{seager2000}
{Seager}, S., \& {Sasselov}, D.~D. 2000, \apj, 537, 916

\bibitem[{{Sing} {et~al.}(2016){Sing}, {Fortney}, {Nikolov}, {Wakeford},
  {Kataria}, {Evans}, {Aigrain}, {Ballester}, {Burrows}, {Deming},
  {D{\'e}sert}, {Gibson}, {Henry}, {Huitson}, {Knutson}, {Lecavelier Des
  Etangs}, {Pont}, {Showman}, {Vidal-Madjar}, {Williamson}, \&
  {Wilson}}]{sing2016}
{Sing}, D.~K., {Fortney}, J.~J., {Nikolov}, N., {et~al.} 2016, \nat, 529, 59

\bibitem[{{Stevenson}(2016)}]{stevenson2016}
{Stevenson}, K.~B. 2016, \apjl, 817, L16

\bibitem[{{Stevenson} {et~al.}(2014{\natexlab{a}}){Stevenson}, {Bean},
  {Fabrycky}, \& {Kreidberg}}]{stevenson2014_gj}
{Stevenson}, K.~B., {Bean}, J.~L., {Fabrycky}, D., \& {Kreidberg}, L.
  2014{\natexlab{a}}, \apj, 796, 32

\bibitem[{{Stevenson} {et~al.}(2014{\natexlab{b}}){Stevenson}, {Bean},
  {Seifahrt}, {D{\'e}sert}, {Madhusudhan}, {Bergmann}, {Kreidberg}, \&
  {Homeier}}]{stevenson2014_w12}
{Stevenson}, K.~B., {Bean}, J.~L., {Seifahrt}, A., {et~al.} 2014{\natexlab{b}},
  \aj, 147, 161

\bibitem[{{Sudarsky} {et~al.}(2003){Sudarsky}, {Burrows}, \&
  {Hubeny}}]{sudarsky2003}
{Sudarsky}, D., {Burrows}, A., \& {Hubeny}, I. 2003, \apj, 588, 1121

\bibitem[{{Thorngren} {et~al.}(2016){Thorngren}, {Fortney}, {Murray-Clay}, \&
  {Lopez}}]{thorngren2016}
{Thorngren}, D.~P., {Fortney}, J.~J., {Murray-Clay}, R.~A., \& {Lopez}, E.~D.
  2016, \apj, 831, 64

\bibitem[{{Tsiaras} {et~al.}(2017){Tsiaras}, {Waldmann}, {Zingales},
  {Rocchetto}, {Morello}, {Damiano}, {Karpouzas}, {Tinetti}, {McKemmish},
  {Tennyson}, \& {Yurchenko}}]{tsiaras2017}
{Tsiaras}, A., {Waldmann}, I.~P., {Zingales}, T., {et~al.} 2017, ArXiv
  e-prints, arXiv:1704.05413

\bibitem[{{Vidal-Madjar} {et~al.}(2011{\natexlab{a}}){Vidal-Madjar}, {Sing},
  {Lecavelier Des Etangs}, {Ferlet}, {D{\'e}sert}, {H{\'e}brard}, {Boisse},
  {Ehrenreich}, \& {Moutou}}]{vidal-madjar2011}
{Vidal-Madjar}, A., {Sing}, D.~K., {Lecavelier Des Etangs}, A., {et~al.}
  2011{\natexlab{a}}, \aap, 527, A110

\bibitem[{{Vidal-Madjar} {et~al.}(2011{\natexlab{b}}){Vidal-Madjar}, {Huitson},
  {Lecavelier Des Etangs}, {Sing}, {Ferlet}, {D{\'e}sert}, {H{\'e}brard},
  {Boisse}, {Ehrenreich}, \& {Moutou}}]{vidal-madjar2011corr}
{Vidal-Madjar}, A., {Huitson}, C.~M., {Lecavelier Des Etangs}, A., {et~al.}
  2011{\natexlab{b}}, \aap, 533, C4

\bibitem[{{Visscher} {et~al.}(2006){Visscher}, {Lodders}, \&
  {Fegley}}]{viisscher2006}
{Visscher}, C., {Lodders}, K., \& {Fegley}, Jr., B. 2006, \apj, 648, 1181

\bibitem[{{Visscher} {et~al.}(2010){Visscher}, {Lodders}, \&
  {Fegley}}]{visscher2010}
---. 2010, \apj, 716, 1060

\bibitem[{{Wakeford} \& {Sing}(2015)}]{wakeford2015}
{Wakeford}, H.~R., \& {Sing}, D.~K. 2015, \aap, 573, A122

\bibitem[{{Wakeford} {et~al.}(2016){Wakeford}, {Sing}, {Evans}, {Deming}, \&
  {Mandell}}]{wakeford2016}
{Wakeford}, H.~R., {Sing}, D.~K., {Evans}, T., {Deming}, D., \& {Mandell}, A.
  2016, \apj, 819, 10

\bibitem[{{Wakeford} {et~al.}(2017{\natexlab{a}}){Wakeford}, {Sing}, {Kataria},
  {Deming}, {Nikolov}, {Lopez}, {Tremblin}, {Amundsen}, {Lewis}, {Mandell},
  {Fortney}, {Knutson}, {Benneke}, \& {Evans}}]{wakeford2017}
{Wakeford}, H.~R., {Sing}, D.~K., {Kataria}, T., {et~al.} 2017{\natexlab{a}},
  Science, 356, 628

\bibitem[{{Wakeford} {et~al.}(2017{\natexlab{b}}){Wakeford}, {Stevenson},
  {Lewis}, {Sing}, {L{\'o}pez-Morales}, {Marley}, {Kataria}, {Mandell},
  {Ballester}, {Barstow}, {Ben-Jaffel}, {Bourrier}, {Buchhave}, {Ehrenreich},
  {Evans}, {Garc{\'{\i}}a Mu{\~n}oz}, {Henry}, {Knutson}, {Lavvas}, {Lecavelier
  des Etangs}, {Nikolov}, \& {Sanz-Forcada}}]{wakeford2017pancet}
{Wakeford}, H.~R., {Stevenson}, K.~B., {Lewis}, N.~K., {et~al.}
  2017{\natexlab{b}}, \apjl, 835, L12

\end{thebibliography}

\appendix

\section{Additional figures and tables}
\counterwithin{figure}{section}
\counterwithin{table}{section}

\begin{figure*}[h!]
\epsscale{0.55}
\plotone{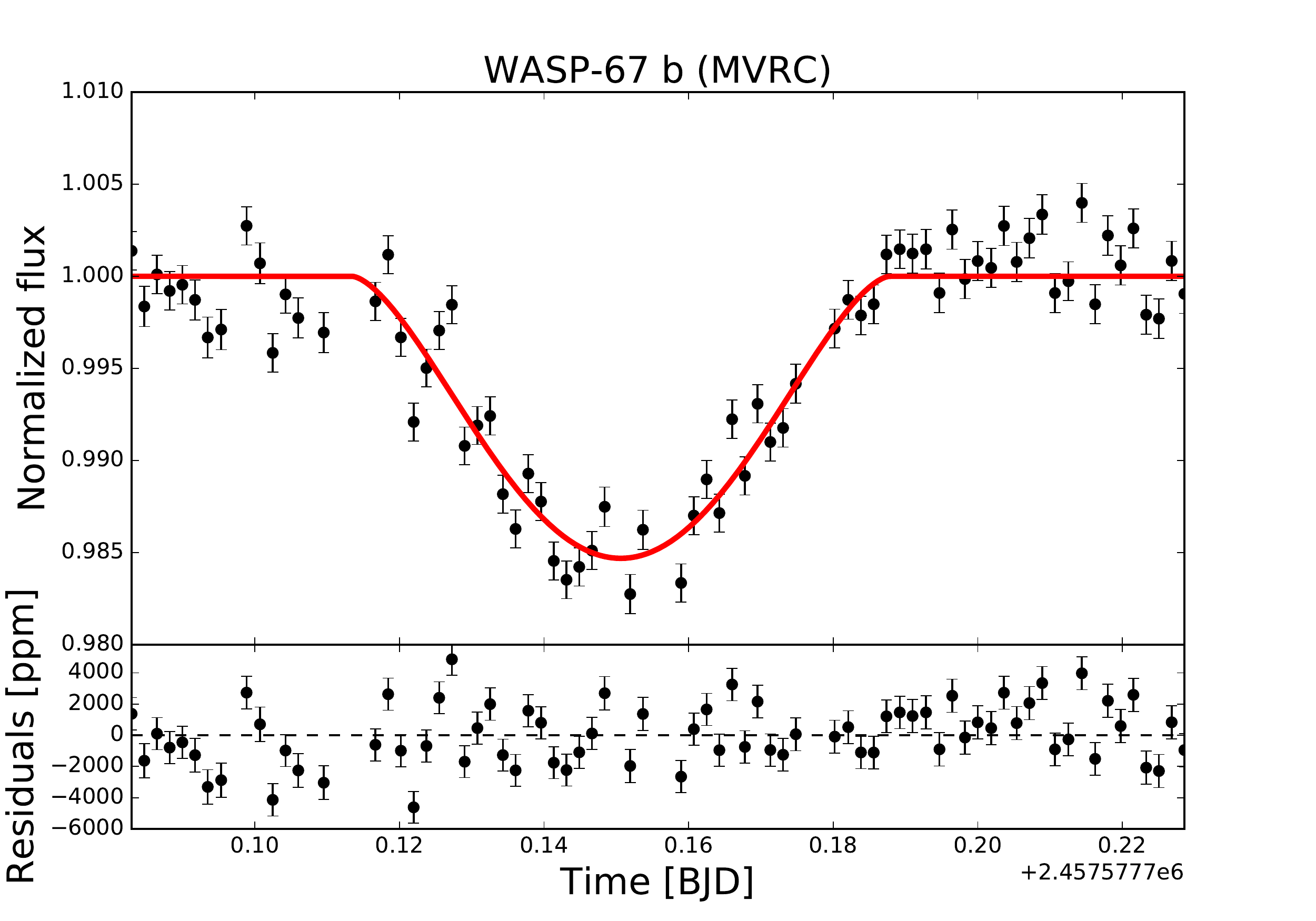}
\epsscale{0.55}
\plotone{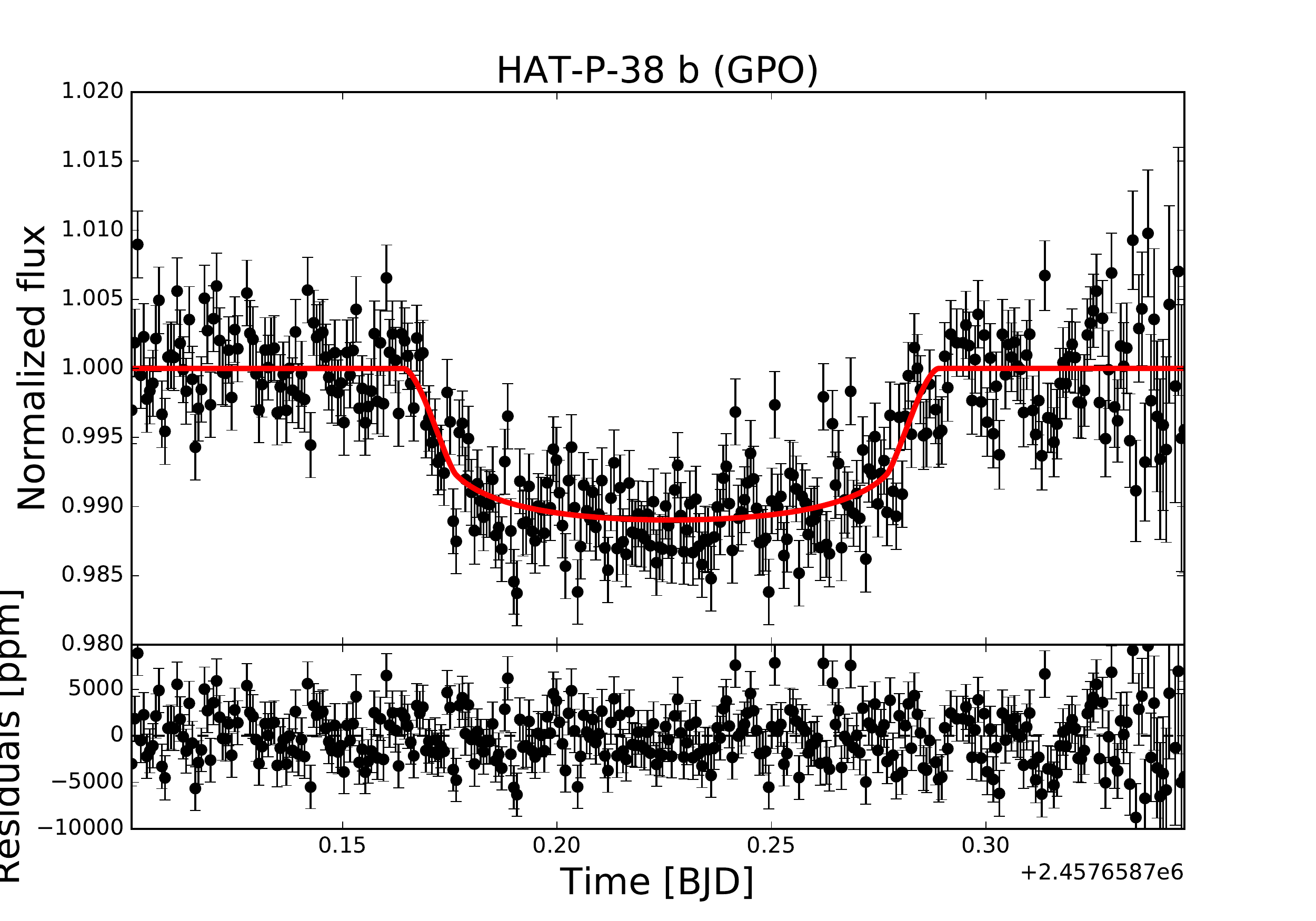}
\caption{MVRC observations of the transit of WASP-67 b ($left$) and GPO observation of the transit of HAT-P-38 b ($right$), with their respective best-likelihood fits.}
\label{transits_kelt}
\end{figure*}

\begin{table}
\begin{center}
\caption{Limb darkening coefficients for WASP-67 (left) and HAT-P-38 (right). The first column of each table indicates the wavelength range corresponding to the channel. The second and third columns are for the linear ($u_a$) and quadratic ($u_b$) coefficients. For each star, the last row denotes the coefficients for the band-integrated light curve.}
\label{limbdtable}
\begin{tabular}{c|c|c}
\hline
$\lambda$ [$\mu$m] &  $u_a$ & $u_b$ \\
\hline
1.125-1.157	& 0.286	& 0.170	\\
1.157-1.188	& 0.274	& 0.176	\\
1.188-1.220	& 0.267	& 0.182	\\
1.220-1.252	& 0.260	& 0.188	\\
1.252-1.284	& 0.244	& 0.203	\\
1.284-1.315	& 0.241	& 0.209	\\
1.315-1.347	& 0.230	& 0.213	\\
1.347-1.379	& 0.218	& 0.226	\\
1.379-1.410	& 0.208	& 0.239	\\
1.410-1.442	& 0.199	& 0.242	\\
1.442-1.474	& 0.187	& 0.250	\\
1.474-1.506	& 0.170	& 0.249	\\
1.506-1.537	& 0.152	& 0.271	\\
1.537-1.569	& 0.135	& 0.274	\\
1.569-1.601	& 0.133	& 0.270	\\
1.601-1.633	& 0.119	& 0.275	\\	
1.125-1.633	& 0.220	& 0.218	\\
\hline
\end{tabular}
\quad
\begin{tabular}{c | c | c }
\hline
$\lambda$ [$\mu$m] &  $u_a$ & $u_b$ \\
\hline
1.125-1.153 &  0.173   & 0.288	\\
1.153-1.181 & 	0.169   & 0.291\\
1.181-1.209 & 	0.162   & 0.295\\
1.209-1.237 & 	0.156   & 0.298\\
1.237-1.265 & 	0.151   & 0.302\\
1.265-1.292 & 	0.131   & 0.312\\
1.292-1.320 & 	0.137   & 0.316\\
1.320-1.348 & 	0.129   & 0.319\\
1.348-1.376 & 	0.121   & 0.324\\
1.376-1.404 & 	0.112   & 0.337\\
1.404-1.432 & 	0.108   & 0.333\\
1.432-1.460 & 	0.098   & 0.338\\
1.460-1.488 & 	0.085   & 0.347\\
1.488-1.516 & 	0.080   & 0.345\\
1.516-1.544 & 	0.066   & 0.353\\
1.544-1.571 & 	0.061   & 0.351\\
1.571-1.599 & 	0.058   & 0.344\\
1.599-1.627 &		0.048   &0.346\\
1.125-1.627 &		0.122   &0.319\\
\hline
\end{tabular}
\end{center}
\end{table}

\begin{table}
\begin{center}
\caption{Fitted radius ratio $k_r$ for the spectroscopic transits of WASP-67 b. The left column corresponds to the central wavelength of each spectral channel.}
\label{radiusratio_w67}
\begin{tabular}{c|c}
\hline
$\lambda$ [$\mu$m] &  WASP-67 b \\
\hline
1.172 & $0.16035 \pm 0.00090 $ \\
1.204 & $0.15914 \pm 0.00081 $ \\
1.236 & $0.16048 \pm 0.00064 $ \\
1.268 & $0.16125 \pm 0.00079 $ \\
1.300  &    $0.15984 \pm 0.00072 $ \\
1.331 & $0.16049 \pm 0.00067 $ \\
1.363 & $0.16181 \pm 0.00069 $ \\
1.394 & $0.16177 \pm 0.00069 $ \\
1.426 & $0.16169 \pm 0.00072 $ \\
1.458 & $0.16054 \pm 0.00070 $ \\
1.490 & $0.15905 \pm 0.00068 $ \\
1.522 & $0.15979 \pm 0.00076 $ \\
1.553 & $0.15985 \pm 0.00092 $ \\
1.585 & $0.16105 \pm 0.00071 $ \\
1.617 & $0.16088 \pm 0.00086 $ \\
\hline
\end{tabular}
\end{center}
\end{table}

\begin{table}
\begin{center}
\caption{Fitted radius ratio $k_r$ for the spectroscopic transits of the first (second column), second (third), and combined (fourth) visit of HAT-P-38 b. The leftmost column corresponds to the central wavelength of each spectral channel.}
\label{radiusratio_h38}
\begin{tabular}{c|c|c|c}
\hline
$\lambda$ [$\mu$m] &  First visit & Second visit & Combined\\
\hline
1.139 & $0.09168 \pm 0.00082 $ & $0.09273 \pm 0.00070 $ & $0.09218 \pm 0.00055 $\\
1.167 & $0.09259 \pm 0.00078 $ & $0.09243 \pm 0.00065 $ & $0.09251 \pm 0.00052 $ \\
1.195 & $0.09096 \pm 0.00062 $ & $0.09268 \pm 0.00071 $ &  $0.09183 \pm 0.00049 $\\
1.223 & $0.09200 \pm 0.00072 $  & $0.09247 \pm 0.00066 $ & $0.09222 \pm 0.00048 $\\
1.251 & $0.09283 \pm 0.00050 $ & $0.09339 \pm 0.0007 $ & $0.09308 \pm 0.00043 $ \\
1.278 & $0.09220 \pm 0.00070 $ & $0.09231 \pm 0.00049 $ & $0.09227 \pm 0.00044 $\\
1.306 & $0.09273 \pm 0.00076 $ & $0.09162 \pm 0.00057 $ & $0.09219 \pm 0.00048 $\\
1.334 & $0.09236 \pm 0.00054 $ & $0.09239 \pm 0.00059 $ & $0.09237 \pm 0.00039 $\\
1.362 & $0.09241 \pm 0.00069 $ & $0.09364 \pm 0.0007 $ & $0.09299 \pm 0.00049 $\\
1.390 & $0.09291 \pm 0.00055 $& $0.09345 \pm 0.00073 $ & $0.09316 \pm 0.00046 $ \\
1.418 & $0.09338 \pm 0.00055 $ & $0.09385 \pm 0.00057 $ &  $0.09359 \pm 0.00039 $\\
1.446 & $0.09324 \pm 0.00070 $ & $0.09428 \pm 0.00069 $ &  $0.09375 \pm 0.00049 $\\
1.474 & $0.09337 \pm 0.00057 $ & $0.09419 \pm 0.00065 $ &  $0.09377 \pm 0.00043 $\\
1.502 & $0.09380 \pm 0.00069 $ & $0.09507 \pm 0.00078 $ &  $0.0944 \pm 0.00053 $\\
1.530 & $0.09351 \pm 0.00068 $ & $0.09362 \pm 0.00067 $ &  $0.09356 \pm 0.00047 $\\
1.558 & $0.09325 \pm 0.00068 $ & $0.09272 \pm 0.00070 $ & $0.09301 \pm 0.00049 $\\
1.585 & $0.09377 \pm 0.00072 $ & $0.09278 \pm 0.00067 $ & $0.0933 \pm 0.00049 $\\
1.613 & $0.09424 \pm 0.00081 $ & $0.09135 \pm 0.00068 $ & $0.09287 \pm 0.00056 $\\
\hline
\end{tabular}
\end{center}
\end{table}

\begin{table}
\begin{center}
\caption{Reduced $\chi^2$ for the spectroscopic channels of the two visits of HAT-P-38 b.}
\label{chisq}
\begin{tabular}{c | c | c}
\hline
$\lambda$ [$\mu$m]  &  First visit   &  Second visit  \\
\hline
1.153-1.181  &  0.9996    &   0.9999    \\
1.125-1.153  &  0.9998    &   0.9995    \\
1.181-1.209  &  0.9996    &   0.9998    \\
1.209-1.237  &  1.0000    &   0.9999    \\
1.237-1.265  &  0.9997    &   0.9999    \\
1.265-1.292  &  0.9997    &   0.9997    \\
1.292-1.320  &  0.9999    &   0.9998    \\
1.320-1.348  &  0.9997    &   0.9988    \\
1.348-1.376  &  0.9997    &   0.9998    \\
1.376-1.404  &  0.9998    &   0.9999    \\
1.404-1.432  &  0.9998    &   0.9999    \\
1.432-1.460  &  0.9998    &   0.9996    \\
1.460-1.488  &  0.9996    &   1.0000    \\
1.488-1.516  &  0.9997    &   0.9997    \\
1.516-1.544  &  0.9999    &   0.9999    \\
1.544-1.571  &  0.9997    &   0.9999    \\
1.571-1.599  &  0.9999    &   0.9998    \\
1.599-1.627  &  0.9998    &   0.9999    \\
\hline
\end{tabular}
\end{center}
\end{table}

\end{document}